%% file: 00_main.tex
\documentclass[manuscript]{acmart} 

\usepackage{graphicx}
 \usepackage{soul}
 \usepackage{color}
   
\usepackage{multirow}

\usepackage{subfig}
\usepackage{longtable}
\usepackage{framed}

\setcopyright{acmlicensed}
\acmJournal{PACMHCI}
\acmYear{2025} \acmVolume{9} \acmNumber{2} 
\acmArticle{CSCW198} \acmMonth{4}\acmDOI{10.1145/3711096}

\newcommand{\pquote}[1]{{\textit{\textcolor{black}{``#1''}}}}

\usepackage{mdframed}
\usepackage{changepage} 
\newenvironment{quotebox}{
  \vspace{5pt} 
  \begin{adjustwidth}{20pt}{20pt} 
  \begin{mdframed}[
    linecolor=black,          
    linewidth=1pt,           
    topline=false,              
    bottomline=false,          
    rightline=false,            
    leftline=true,               
    backgroundcolor=white,
    innerrightmargin=0pt,       
  ]
  \itshape 
}{
  \end{mdframed}
  \vspace{5pt} 
  \end{adjustwidth}
}

\begin{document}
\title{Understanding the Challenges of Maker Entrepreneurship}

\author{Natalie Friedman}
\authornote{Contact author}
\email{nvfriedm@gmail.com}
\affiliation{%
  \institution{Cornell Tech}
  \streetaddress{2 W Loop Rd}
  \city{New York}
  \state{New York}
  \country{USA}
  \postcode{10044}
}

\author{Alexandra Bremers}
\email{awb227@cornell.edu}
\affiliation{%
  \institution{Cornell Tech}
  \streetaddress{2 W Loop Rd}
  \city{New York}
  \state{New York}
  \country{USA}
  \postcode{10044}
}

\author{Adelaide Nyanyo}
\email{aen45@cornell.edu}
\affiliation{%
  \institution{Cornell Tech}
  \streetaddress{2 W Loop Rd}
  \city{New York}
  \state{New York}
  \country{USA}
  \postcode{10044}
}

\author{Ian Clark}
\email{iclark@andrew.cmu.edu}
\affiliation{%
  \institution{HCI Institute, Carnegie Mellon University}
  \streetaddress{5000 Forbes Ave.}
  \city{Pittsburgh}
  \state{Pennsylvania}
  \country{USA}
  \postcode{15213}
}

\author{Yasmine Kotturi}
\email{kotturi@umbc.edu}
\affiliation{%
  \institution{University of Maryland, Baltimore County}
  \streetaddress{1000 Hilltop Circle}
  \city{Baltimore}
  \state{Maryland}
  \country{USA}
  \postcode{21250}
}

\author{Laura Dabbish}
\email{dabbish@cmu.edu}
\affiliation{%
  \institution{HCI Institute, Carnegie Mellon University}
  \streetaddress{5000 Forbes Ave.}
  \city{Pittsburgh}
  \state{Pennsylvania}
  \country{USA}
  \postcode{15213}
}

\author{Wendy Ju}
\email{wendyju@cornell.edu}
\affiliation{%
  \institution{Cornell Tech}
  \streetaddress{2 W Loop Rd}
  \city{New York}
  \state{New York}
  \country{USA}
  \postcode{10044}
}

\author{Nikolas Martelaro}
\email{nikmart@cmu.edu}
\affiliation{%
  \institution{HCI Institute, Carnegie Mellon University}
  \streetaddress{5000 Forbes Ave.}
  \city{Pittsburgh}
  \state{Pennsylvania}
  \country{USA}
  \postcode{15213}
}

\renewcommand{\shortauthors}{Natalie Friedman et al.}

\begin{abstract}
The maker movement embodies a resurgence in DIY creation, merging physical craftsmanship and arts with digital technology support. However, mere technological skills and creativity are insufficient for economically and psychologically sustainable practice. By illuminating and smoothing the path from ``maker" to ``maker entrepreneur," we can help broaden the viability of making as a livelihood. Our research centers on makers who design, produce, and sell physical goods. In this work, we explore the transition to entrepreneurship for these makers and how technology can facilitate this transition online and offline. We present results from interviews with 20 USA-based maker entrepreneurs {(i.e., lamps, stickers)}, six creative service entrepreneurs {(i.e., photographers, fabrication)}, and seven support personnel (i.e., art curator, incubator director). Our findings reveal that many maker entrepreneurs 1) are makers first and entrepreneurs second; 2) struggle with business logistics and learn business skills as they go; and 3) are motivated by non-monetary values. We discuss training and technology-based design implications and opportunities for addressing challenges in developing economically sustainable businesses around making.

\end{abstract}


\maketitle

\input{01_introduction}

\input{01b_relatedwork}

\input{02_method}

\input{03_results}

\input{04_discussion}
\input{05_conclusion}

\begin{acks}
We thank all the research participants for their participation in our study.
We thank Rachel Liao and Kathy Song who participated in in this project.
Quentin Romero Lauro and Courtney Kreitzer who participated in a summer REU related to this research.
This material is based upon work supported by the National Science Foundation under Grant No. 2222719.
\end{acks}

\bibliographystyle{ACM-Reference-Format}

\bibliography{00_main}

\received{July 2024}
\received[revised]{October 2024}
\received[accepted]{December 2024}

\input{Appendix}

\end{document}

%% file: 01_introduction.tex
\section{Introduction}
\label{sec:intro}
The "Maker Movement" champions the do-it-yourself creation of physical products, crafts, and arts, with underlying support from technologies to enable the creation of designs and fabrication of goods~\cite{anderson_makers_2012}. 
Many makers who make physical goods seek to turn their creative endeavors into business opportunities by selling their products independently. 
This requires makers to not only learn the tools and processes associated with their craft, but also additional skills to transform their work into a sustainable livelihood~\cite{hui2017makerspaces-entrepreneurship}. 
For these ``maker entrepreneurs''---sole proprietors who actively attempt to commercialize their {physical} creative work ~\cite{anderson_makers_2012}---the tools they turn to for promotion, production, and distribution are often digital~\cite{duffy2017gender,sahut2021age,troxler2017digital}, such as {social media platforms}~\cite{kotturi2021unique, browder_emergence_2019}, {computer-aided design software and digital fabrication tools (i.e., laser cutters)} ~\cite{anderson_makers_2012, lang_zero_2013, rosner2014making}, {and online platforms and eCommerce websites} ~\cite{troxler2017digital}.
However, despite the abundance of digital tools available for people to run businesses on their own, many find the transition to entrepreneurship and effective use of such digital tools challenging due to a lack of entrepreneurial knowledge \cite{dillahunt2018entrepreneurship}.
{While past works have explored micro-entrepreneurship, many have focused on transitioning current entrepreneurs towards digital business practices} \cite{dillahunt2018entrepreneurship} {and exploring digital tool use for entrepreneurs in low-resourced and rural contexts} \cite{afroze2014women,meera2022innovative,sultan2020exploratory}.
{There are only a few examples of research that focus specifically on maker entrepreneurs} \cite{hui2017makerspaces-entrepreneurship,homepreneurship} {whose creative goals influence their business goals, and that suggest design opportunities for new digital tools to support these entrepreneurs} \cite{kotturi2024peerdea}.
{This suggests that more research is needed at the intersection of makers, entrepreneurship, and technology to understand better how makers, who in theory have access to a plethora of technologies, undergo the transition to entrepreneurship and how technologies may be designed to facilitate this transition}~\cite{browder_emergence_2019,duffy2018, kotturi2021unique}.

The present research builds on human-computer interaction (HCI) and computer-supported cooperative work (CSCW) research in micro-entrepreneurship \cite{10.1145/3411764.3445126, hui2017makerspaces-entrepreneurship, socialcapital, 10.1145/3411764.3445146} to focus specifically on the entrepreneurship of ``makers,'' who, at their core, conceptualize, design, and produce physical goods and sell them ~\cite{aldrich_democratization_2014}. 
We argue that this focus on maker businesses warrants special attention because of the unique difficulties that can arise due to the physical and small-scale nature of these businesses, such as acquiring specialty equipment to make products, sourcing materials at small scale, storing and shipping products, and more~\cite{anderson_makers_2012}. 
For this research initiative, we work with a broader (and more inclusive) definition of makers suggested by ~\cite{dougherty_maker_2012} that considers makers to be anyone who is an enthusiast of creating physical goods.
Our {motivation is to suggest opportunities for new technological platforms, tools, and education to support makers in running successful businesses.}

Based on the related research {around micro-entrepreneurship and the use of technology to facilitate small entrepreneurial ventures}, we focus on the following research questions:

\begin{itemize}

\item \textbf{RQ1:} What is involved in the transition from maker to maker entrepreneur?

\item \textbf{RQ2:} What are business challenges unique to maker entrepreneurs?

\item \textbf{RQ3:} What role can technology play in supporting maker entrepreneur business activities?

\end{itemize}

To address these three research questions, we interviewed 26 {US-based} maker entrepreneurs who are {part of the creative economy} as defined by Khaire ~\cite{khaire2021entrepreneurship} {as people who create cultural objects.} These interviewees included 20 maker entrepreneurs, {who create and sell physical goods}, and six people who provide creative services. 
To complement entrepreneurs' perspectives, we also interviewed seven support personnel: intermediaries providing consulting, mentorship, or business services for the creative entrepreneurship community {and who are knowledgeable about the process makers take in becoming entrepreneurs. These interviews were conducted on Zoom in 2023, as COVID-19 was impacting our ability to run in-person interviews.}

We qualitatively analyzed these interviews to inform the future design of tools to support maker entrepreneurship.
Our findings suggest that maker entrepreneurs focus on their making first and entrepreneurship second, {meaning that they prioritize the art of making, process, and aesthetics before focusing on financial management. In fact, many entrepreneurs discussed an ``accidental'' entrance into entrepreneurship and grappled with business basics, like keeping receipts or records of production costs.} 
We also find that maker entrepreneurs prioritize values such as the authenticity of {relationships with their customers} and community {building with both customers and manufacturers}, influencing their business decision-making. 
{For instance, participants expressed how they were not always interested in financially optimizing their business and instead preferred to focus on other metrics of business success, like relationships and strong creative identity.
However, we also find that such non-monetary values are hard to track with traditional business software.} 
Finally, we find that to make up for limited business training, the maker entrepreneurs in our study learned to run their businesses through trial-and-error, informal and {accessible resources, like online videos, the library, the internet}, and some ongoing mentorship {from people like, an entrepreneurship professor or art incubator director.
Such practices are often created by makers independently, suggesting a need for more educational support for makers turning to entrepreneurship.}

{Overall, we contribute a descriptive view on the state of maker entrepreneurs' common journey and challenges with balancing the financial, logistical, and creative aspects of being a maker entrepreneur.
We suggest unmet needs that computer support could address to help makers become effective entrepreneurs, adding to the CSCW literature on technology support for micro-entrepreneurs in general.
Finally, we conclude with possible design guidelines for business operations and learning technologies to help makers throughout their journey to successful and fulfilling entrepreneurship.}

%% file: 01b_relatedwork.tex
\section{Related Work}
\label{Sec:RelatedWorkb}

We draw on three bodies of scholarship to inform our study of maker entrepreneurs and the tools that could support them. These bodies of literature are (1) maker and creative entrepreneurship, (2) entrepreneurship in human-computer interaction (HCI), and (3) computing technologies for maker entrepreneurs.
 
\subsection{What defines maker entrepreneurs?}
\subsubsection{Common definitions}
According to business administration researchers Troxler \& Wolf, ``maker entrepreneurs'' are a subset of ``creative entrepreneurs'' \cite{troxler2017digital} who work on a tangible product (i.e., woodcrafts, jewelry, ceramics) rather than provide a service (i.e., photography, graphic design, video production) \cite{deresiewicz2015death}. 
Maker entrepreneurs generate revenue from the design, production, and sale of products---they are, therefore, explicitly not hobbyists who practice a craft for enjoyment alone ~\cite{wolf2017maker}. 

Maker entrepreneurs are often motivated to generate revenue for different reasons, such as generating income as a full-time business, as a source of secondary income, or as a way to sustain their creative practice \cite{williams2021genz,miller_etsy_2007}.
For example, \citet{homepreneurship} profiled ``homepreneurs'' who used desktop cutting plotters to personalize and sell goods online, finding that 65\% of the 49 surveyed homepreneurs started their businesses out of their crafting hobby.
Only 8\% surveyed stated that their craft work was a full-time job; 22\% considered it to be part-time, 45\% thought it was a hobby, and 24\% viewed it as both a hobby and a part-time job. 
These varying motivations and business configurations influence how maker entrepreneurs view themselves and how they operate their businesses.

\subsubsection{Balancing maker and business identities}
\label{sec:balancingMB}
Maker entrepreneurs encounter a dilemma: On the one hand, they create products in an often artisanal manner---products with aesthetics and cultural value beyond pure utility \cite{pret2019artisan}---that may reflect non-materialistic, counter-cultural, and non-conformist values \cite{solomon2020artisans}. These products also demonstrate symbolic value, which \citet{tan2003leveraging} described as \textit{``intangible benefits obtained only when the receiving person understands and shares the same meanings as the person who gives it. Symbolic meanings are derived through the socialization process where individuals learn to agree on shared meanings of some symbols or objects, and also to develop individual symbolic interpretations of their own.''} On the other hand, these products are meant for sale, a fact which introduces makers to customers, production schedules, deadlines, negotiations, and other aspects of commerce that may seem antithetical to makers' self-expression and identity \cite{defillippi_introduction_2007, eikhof_for_2007, michlewski_uncovering_2008, gotsi_managing_2010, glaveanu_creativity_2014}. 
Conflicting artistic and economic logic in creative work belie a tension between makers’ existing creative identity and their burgeoning identity as an entrepreneur, a tension that may complicate makers’ transition to a new role that merges their identities \cite{kreiner_where_2006, gotsi_managing_2010, glaveanu_creativity_2014}. Yet this transition is critical because as they shift to maker entrepreneurship, makers must manage a wide variety of business functions---product design, manufacturing, marketing, sales, finance, and human resources---while still sustaining the artistic sensibility that engendered the product in the first place \cite{doussard_manufacturing_2018}.
Makers who seek to create their own businesses but who fail to merge their dissonant identities may struggle. 
For example, they may commit themselves to creative growth given their desire to make art but feel unsure of how to pursue business growth in a way that aligns with their creative values.
For those who view making as an antidote to or statement against mass production, the turn to entrepreneurship may contradict an alternative, often rebellious mindset \cite{banks_autonomy_2010,luckman_aura_2013}. 

Similarly, as the demand for entrepreneurial tasks increases, creative tasks are subsequently deprioritized. 
This realistic consequence of sole proprietorship may leave budding maker entrepreneurs grappling with a loss of creative identity and energy \cite{mcrobbie_clubs_2002}. Makers who strive towards entrepreneurial goals but who fail to support themselves with their work experience psychological stress and grief, especially if they must turn to non-creative work to support themselves \cite{hennekam_involuntary_2016, duffy2017not}.
Understanding how makers transition to becoming entrepreneurs today can help inform ways to support them with future tools and systems.

\subsubsection{Transitioning from making into maker entrepreneurship}

Interviews with makers who have successfully made the leap to maker entrepreneurship suggest that makers may need to go through stages of identity development \cite{werthes_cultural_2018, michlewski_uncovering_2008, poorsoltan_artists_2012, berglund_opportunities_2020}. 
Initially, they may reject an entrepreneurial identity as they pursue their creativity, eventually shaping entrepreneurialism on their own terms (e.g., not devoted to commerce or consumption) through self-reflection and interaction with other maker entrepreneurs \cite{werthes_cultural_2018}. 
Such a reconsideration of the self as a maker to a maker entrepreneur is often possible as makers often share many of the same characteristics as entrepreneurs, including creativity, desire to innovate, tolerance for ambiguity, risk-taking, and an internal locus of control \cite{michlewski_uncovering_2008, poorsoltan_artists_2012}.
Moreover, recent scholarship suggests that entrepreneurship itself is best conceptualized as artifact-centered design, which would presumably position it firmly in makers’ domain \cite{berglund_opportunities_2020}. 
To successfully make the transition to entrepreneurship, makers may require guidance in the form of coaching, workshops, and other mechanisms that, in the language of social identity theory \cite{tajfel_integrative_1979}, help makers to stop casting entrepreneurs as the out-group and begin seeing themselves in the \textit{maker entrepreneur} in-group \cite{werthes_cultural_2018}, thus allowing them the break their sense as a non-entrepreneur, recognize the characteristics that can make them an entrepreneur, and rebuild their own self-image as both a creative and entrepreneur.
Such support mechanisms may prove particularly helpful for women and underrepresented minority makers who often have difficulty reconciling the demands of professional identities that they view as existing in conflict with each other \cite{tracy_fracturing_2006}.

\subsubsection{Variances by culture and context}

The physical nature of maker businesses brings challenges that other creative, independent workers, such as musicians or graphic designers, do not have \cite{anderson_makers_2012}.
Early on, maker entrepreneurs must make the products, physically manage and store their inventory, and ship their products by themselves, on top of the other roles of sourcing reliable materials and managing finances \cite{doussard_manufacturing_2018}. 
On the other hand, maker entrepreneurs are often able to take advantage of their small scale; limited startup costs make it possible for some entrepreneurs to start their businesses without external capital, and they can use small batch manufacturing to create limited editions and respond rapidly to trends~\cite{lipson2010factory}.
While many of these core aspects of operating a maker business are constant, differences exist among cultures and contexts, and an understanding of maker entrepreneurship across them can inform a broader understanding of how makers become entrepreneurs.

{CSCW researchers} ~\citet{DIY,DIY2} {have also investigated why people from lower socio-economic backgrounds are motivated to use Do-It-Yourself (DIY) methods and found two primary reasons: sustainable \& economical living and social \& community wellbeing. However, these participants did not discuss businesses past their own personal DIY practices.}

Scholars have detailed maker entrepreneur pursuits in various geographical contexts (e.g., Europe~\cite{bhansing2018passion}, China~\cite{lindtner2015designed, lindtner2015hacking}, India~\cite{meera2022innovative}, and Bangladesh~\cite{sultan2020exploratory,afroze2014women}), highlighting that the issues facing maker entrepreneurs are often specific to local cultures and economies.
For instance, through five years of ethnographic work throughout urban regions of China, Lindtner~\cite{lindtner2015hacking} details how China's expertise in manufacturing, repair, and reuse provides a unique foundation for implementing the vision of making beyond Western ideals of openness---Chinese makers do so out of necessity due to the dearth of mass-manufactured products in the region and the culture of creative reuse and recomposition of component technologies into new products.
Sultana et al.~\cite{sultan2020exploratory} and Afroze et al.~\cite{afroze2014women} have explored the ways that the caste system and patriarchal norms have shaped the dynamics of artisan craft entrepreneurship in India and Bangladesh, respectively. Their findings show how many women often lack entrepreneurial training, are discriminated against, and have limited shared entrepreneurial knowledge; however, through perseverance, support from their family, and increasing social media savviness, they can overcome these challenges to form successful businesses. {Similar findings on challenges were established by CSCW researchers} \citet{womenIntersectional} {who studied challenges faced by women of color in nine maker spaces in Australia. They found that these challenges similarly included first impressions and visual representations, dimensions of the enabling environment, role of community partnerships, and intersectional identities of women. We start to see a pattern in CSCW literature on challenges and methods to overcome entrepreneurial hardships, as} \citet{lee2023refugee} {found ways similar to} \citet{afroze2014women} {that refugees overcome challenges in entrepreneurship: support network, self-efficacy, and the growth mindset.} 

Howard et al.~\cite{howard2014maker} describe the European Maker Movement as highly driven by the desire to compete with the US and emerging economies. 
Unlike the US, where the Maker Movement has been encouraged by Silicon Valley and its lack of previous industrial structure, European makers often operate from a strong heritage of artisanship but with less cultural tolerance for personal financial risk \cite{howard2014maker}. 
Essig \cite{essig2017same} further describes the economic differences among European and US maker movements, suggesting that discussions of arts and cultural entrepreneurship in the EU focus more on helping cultural industries be more economically self-sustaining and less reliant on state funding, whereas in the US ``arts entrepreneurship'' focuses on artist self-sufficiency and career self-management.
In this paper, we focus on US maker entrepreneurship due to the focus on self-sufficiency and independence, noting the perspective of US-centric capitalist frameworks of business success and historical ideals around independent business development and self-determination that may differ from other cultures and contexts. 
Furthermore, we also consider the specific technology context of the US, as described in the next section.

\subsection{How do technological developments affect maker entrepreneurship?}
\subsubsection{Digital platforms affect paths to entrepreneurship}

The emergence of mobile-compatible digital marketplaces \cite{zhang_cold_2017,acker_venmo_2018,etsy_sell_nodate} allow makers to create, market, and sell their goods, without the overhead of running a brick-and-mortar shop \cite{pop2019online}. 
Such platforms allow maker entrepreneurs to grow their businesses well beyond their local area and have attracted many people to start new ventures, especially among those with increased access to small-scale fabrication equipment (i.e., 3D printers, laser cutters)~\cite[Chapter 2]{von2016free}.
For example, \citet{shultz_work_2015} describes that people are acting as their own intermediaries through social media, as opposed to traditional curators. While this could even out inequities since makers no longer have to find curators, networks still matter, and this is less time that they spend on their craft. 

As evidence of the growth in maker entrepreneurship, the number of Etsy sellers more than doubled between 2019 and 2021 to 5.3 million~\cite{etsytransparency2019, etsy2021}.
Many of these new businesses were started by women and other minority entrepreneurs~\cite{etsywomen}.
However, while digital maker store platforms such as Etsy may reduce certain barriers to entry for some, becoming digitally engaged as an entrepreneur is far from a simple pursuit. 
Digital tools and platforms can pose challenges for maker entrepreneurs, such as alienating them from the entrepreneur and creative community that might otherwise be had in local communities~\cite{garvin2023counter}.
Further, digital literacy barriers can heighten inequities given the need to learn digital and algorithmic skills for effective tool use~\cite{dillahunt2018entrepreneurship}, especially in the context of people who prefer to work with their hands to produce physical goods~\cite[Chapter 5]{klawitter2017independent}.
Addressing such challenges is an area for research and design at the intersection of human-computer interaction and small-scale entrepreneurship.

\subsubsection{Entrepreneurship in HCI literature}

Human-computer interaction (HCI) scholarship has explored how entrepreneurs engage their businesses digitally through online marketplaces~\cite{dillahunt2018entrepreneurship}, crowdfunding~\cite{gerber2013crowdfunding}, social media~\cite{kotturi2021unique, israni2023opportunities, dabbish2012social}, automation and AI-assisted technologies~\cite{garvin2023counter}, and online communities~\cite{hui2019distributed}. 
HCI scholarship has maintained an expansive notion of entrepreneurship which has focused on the inequities among entrepreneurs due to uneven access to digital resources and differing levels of digital literacy \cite{dillahunt2018entrepreneurship,hui2018making}. 
In focusing on entrepreneurs from ``lean economies,'' HCI scholars have detailed the behind-the-scenes work that entrepreneurs with fewer resources employ to become entrepreneurs in a digital era~\cite{avle2019additional}.

For instance, many have highlighted the importance of offline social support when acquiring entrepreneurial and technology skills~\cite{casey2014critical, hui2020community, dillahunt2022village, kotturi2022tech, hui2017makerspaces-entrepreneurship, lindtner2014emerging}. 
Forming offline social support groups can make it easier for entrepreneurs to learn how their peers use technology for their business and exchange trustworthy advice~\cite{hui2020community}. 
More specifically, offline social support offers intention and authenticity among entrepreneurs, which can make it easier to vet peers as compared to online interactions ~\cite{kuhn2016near}.
However, it remains unclear whether and how the challenges of digital entrepreneurial tools apply to maker entrepreneurs in particular. 

Therefore, we extend the research in HCI on entrepreneurship by detailing how maker entrepreneurs use digital tools and engage in online and offline social networks to provide support for each others' tool discovery and tool use.

\subsection{Research gap: business support technologies for maker entrepreneurs}

\textit{{``Digital technologies play a central role in bridging access to resources, connections, and opportunity'' (pg. 1)}} \cite{avle_how_2017}
Entrepreneurship is heavily mediated by technology, thus CSCW research \cite{homepreneurship, kotturi2024peerdea} has been concerned with understanding who these people are and learning how they achieve their goals. Many of the work has focused on people in rural and under-resourced environments. 

Much of the work in {CSCW} and HCI has focused on understanding their populations, as well as a few works that developed new tools \cite{kotturi2024peerdea}. Our work focuses on creative entrepreneurship, since this is lightly discussed. Creative entrepreneurship has similarities to micro-entrepreneurship, but we are emphasizing the differences. 

Tools that are designed to help entrepreneurs have been critiqued to disproportionately favor people who have entrepreneurial advantages \cite{dillahunt2018entrepreneurship}.
To date, prior technology research in the creative entrepreneurship space has overwhelmingly focused on creativity support tools for those engaged in open-ended creative work, {but this research does not cover creative support tools with business-oriented functions (i.e., finances, operations, marketing).} Instead, the creative support tools have features like brainstorming ~\cite{foong2017online}, getting feedback on creations~\cite{kotturi2021unique, kotturi2024peerdea}, self-presentation \cite{crain2017share,xu2015classroom,dabbish2012social,kim2017mosaic}, and setting open-ended goals~\cite{krishna2021plan}. 

Research on creativity tools often does not touch on the business-oriented functions where makers may have the least training and knowledge. This poses a challenge for budding maker entrepreneurs, who often work in isolation of supportive and knowledge-sharing communities and have difficulties identifying and remedying gaps in their business knowledge \cite{shultz_work_2015}.

Even makers with formal arts and design training face an entrepreneurial training gap; a Strategic National Arts Alumni Project survey reports {that 71\% of respondents felt that entrepreneurial skills were essential to their creative careers, but only 28\% had developed this training in school. 
This discrepancy is alarming, as a majority of art and design school graduates have been self-employed, and a significant portion have started their own enterprises} \cite{skaggs2017strategic}. 
Makers increasingly seek professional development through artist service organizations, speaking to the need for such business education to be adapted to the needs of makers \cite{olshan_after_2017}. 

While commercial systems exist for entrepreneur business support, such as financial software (e.g., Quickbooks, TurboTax), these systems usually assume an operating size and budget not within the scope of individual maker entrepreneurs who are operating at a small scale \cite{khaire_culture_2017}.
Instead, numerous services catering to small-scale production and sales, such as Etsy, Kickstarter, and Square, are available and used by maker entrepreneurs.
However, due to the aforementioned digital skills gap and the challenges of physical good production, even these tools may prove challenging for maker entrepreneurs.
Therefore, we extend the HCI research on entrepreneurship to consider tool support for small-scale makers by detailing the challenges that maker entrepreneurs specifically face when using technologies to support their business practices, as well as providing design implications for technologies to support their business practices.

%% file: 02_method.tex
\section{Methods}
\label{Sec:Methods}

\subsection{Interview Study Design}
We conducted semi-structured interviews focused on challenges, social support systems and communities, personal maker and entrepreneurial identity, and technologies to support business activities. We used semi-structured interview data and a qualitative thematic analysis approach. The semi-structured approach allowed us to use themes to steer the interview rather than relying solely on specific questions \cite{ljungblad2023applying}. The interview questions were developed with human-computer interaction researchers and a specialist in creative entrepreneurship, who had closely studied the value of community in their multi-year, community-based research with a local makerspace focused on equity in tech and entrepreneurship. {In addition, our design drew from challenges described in CSCW literature, as discussed above about identity} \cite{lee2023refugee} {and business logistics} \cite{shultz_work_2015}. {Our identity questions centered on their personal and artistic stories while our business logistic questions focused on business promotion and production.} Interviews were conducted by five trained human-computer interaction researchers.

\subsubsection{Interview Questions}

The \textit{introduction and background questions} asked participants to introduce themselves, describe what motivated them, and describe how they got started, practically and financially. By understanding their motivations and how they began, we focus on their values and goals and aim to understand their personal context. The \textit{business promotion questions} concerned reaching customers and targeting what differentiates the maker as a business. {For example, we asked questions like, ``How do you promote your business?'', ``What selling platforms do you use?''}. The \textit{factory \& production questions} focused on manufacturing products, choices about outsourcing, and pricing. {Here we asked questions like, ``How do you produce your products?'', and ``What challenges have you encountered in manufacturing your products?''}
The \textit{artistic story questions} concerned maker entrepreneurs' personal identity and whether or not they share their identity or life through the products and artifacts they create. Finally, the \textit{reflection questions} gave interviewees a place to consider their challenges, knowledge, and successes. See \nameref{appendixA} for the full list of questions.

\subsection{Participants}
\input{altParticipantTable}

We cast a wide net around creative entrepreneurship to find people who would be eligible for the interviews. This included people who supported creative entrepreneurs, like curators, maker space co-founders, and people with financial expertise.
We recruited potential participants using a snowball sampling method \cite{naderifar2017snowball}. 
Initially, we compiled a list of individuals and creative communities within the U.S. within the researcher's networks which spanned urban and suburban communities and various socio-economic levels given existing partnerships between researchers and local community organizations. Potential candidates were approached through e-mail.

After an initial round of interviews and analysis, we excluded {makers} who created non-creative goods, such as one entrepreneur who made nutritional spice mixtures, or entrepreneurs who provided solely creative services {but not physical goods}, such as masseuses. We decided to do this because we were trying to pinpoint the physical nature of creative making.

Ultimately, we interviewed 33 individuals: 20 maker entrepreneurs whom we define as individuals who make and sell physical items, six creative service entrepreneurs, such as photographers who have some artistic practice as part of their business and thus have similar attributes as maker entrepreneurs, and seven people who support creative entrepreneurs through different kinds of mentorship or business services and have knowledge of maker entrepreneurial practices such as curators and people who operate maker spaces, entrepreneurial hubs, or business consultancy services. We have labeled these ``I,'' for ``intermediaries,'' as defined by Khaire \cite{khaire_culture_2017} as people who are not selling cultural goods directly, but help to build value through cultural commentary. {We required that these intermediaries supported creatives that specifically made physical objects}.
Table \ref{tab:participants} shows a listing of our interview participants with the type of creative good or service they provide, their background, basic demographics, and years of experience. 

\subsection{Procedure}

The information about our study and consent form was shared via email for participants to review and sign. All interviews were conducted over Zoom video conference (n=31) or in person (n=2), during a mutually agreed upon time. 
The interviews lasted between 20 minutes to 83 minutes (\textit{M} = 46 minutes, \textit{SD} = 14 minutes). 
A researcher began the interview using the general questions listed in Appendix \ref{appendixA} and asked follow-up questions as relevant. 
All interviews were recorded and transcribed using an automated speech transcription system, Whisper \cite{rao2023transcribing}. 
Errors in transcriptions were corrected by researchers during the subsequent analysis.

\subsection{Analysis}

Five researchers reviewed and coded the initial set of interviews to {familiarize themselves with the data}, surface comments and develop a list of initial themes related to the transition to entrepreneurship by clustering them through affinity mapping \cite{harboe2015real}. {They discussed their interpretations of each theme. When there was a disagreement about a piece of data belonging to a category, researchers reached a collective agreement through group conversation. They developed mutually-agreed-upon definitions of each theme based on the excerpts used to develop that theme.} 

Six initial themes were formed based on the data: 1) Logistics, 2) Alternative Currencies, 3) Relationships, 4) Identity, 5) Technology, and 6) Mentoring. 
From this point, two team members coded all the interviews based on these six themes, including the interviews used to develop the themes. Because the same coders developed the themes and applied them to all the data, we do not report inter-rater reliability per guidance from \citet{mcdonald2019reliability}. {The interviews were reviewed in total two times by these two coders.}
Coded statements were collected in a spreadsheet and the team met weekly to discuss what was emerging from the data. {This process allowed for continuous refinement and validation of the themes.
Once all interviews were coded, researchers reviewed and discussed each theme's data to further understand the nature of each category.
The collected themes and representative quotes were then used to develop the results and discussion of this paper.}

\subsection{Positionality}
Our team consists of HCI researchers with diverse exposure to creative entrepreneurship and platform-based work. Six team members have connections to hackerspaces or makerspaces, which facilitated participant recruiting and brought a personal connection to some interviewees. One team member with prior experience at Etsy contributed insights into creative entrepreneurship and the maker community, while another team member’s background in economics helped us analyze pricing and expense management strategies. Additionally, a team member’s direct experience as a maker entrepreneur and two members with siblings in creative fields brought further personal perspectives, informing our study's protocol development and analysis.

We organized our analysis process to encourage discussion and consensus, but the inherent subjectivity and interpretive nature of our methodology means that another set of researchers could arrive at different conclusions based on the same interviews.

%% file: altParticipantTable.tex
\begin{footnotesize}
\begin{longtable}{lc p{1.8cm} p{2.5cm}ll p{2.5cm}}
\caption{Maker entrepreneur (P) and Intermediaries (I) interviewees. Maker entrepreneur creative goods are in red text, and creative entrepreneur services are in blue.} \label{tab:participants}\\
\multicolumn{2}{l}{\textbf{ID}} &
  \multirow{2}{=}{\textbf{\color[HTML]{980000}Product / \color{ACMDarkBlue} Service}} &
  \textbf{Background} &
  \textbf{Craft exp.} &
  \textbf{Bus. exp.} &
  \textbf{Other job} \\
  \multicolumn{2}{l}{(Age, Gender)} & & & & 
\endfirsthead
 \caption[]{Maker entrepreneur, creative entrepreneur, and intermediary participants \textit{(continued)}}\\

\multicolumn{2}{l}{\textbf{ID}} &
  \multirow{2}{=}{\textbf{\color[HTML]{980000} Product / \color{ACMDarkBlue} Service}} &
  \textbf{Background} &
  \textbf{Craft exp.} &
  \textbf{Bus. exp.} &
  \textbf{Other job} \\
  \multicolumn{2}{l}{(Age, Gender)} & & & & \endhead \hline
\multicolumn{2}{l}{\textbf{P01} (30-40, M)} &
  {\color[HTML]{980000} Digital Art} &
  \multirow[c]{2}{=}{Woodworking} &
  \multirow[c]{2}{*}{1 - 5} &
  \multirow[c]{2}{*}{1 - 5} &
   \multirow[c]{2}{=}{N/A} \\*
   \\ \hline
\multicolumn{2}{l}{\textbf{P02} (30-40, M)} &
  \multirow[c]{2}{=}{{\color[HTML]{980000} Painting}} &
  \multirow[c]{2}{=}{Musician, Painter, Professor} &
  \multirow{2}{*}{15 - 20} &
  \multirow{2}{*}{1 - 5} &
  \multirow{2}{=}{Professor} \\*
   &
   &
   &
   &
   \\ \hline
\multicolumn{2}{l}{\textbf{P03} (18-30, F)} &
  \multirow[c]{2}{=}{{\color[HTML]{980000} Painting}} &
  \multirow{2}{=}{Business + Fashion degree} &
  \multirow{2}{*}{1 - 5} &
  \multirow{2}{*}{1 - 5} &
  \multirow{2}{=}{Waitress, tattoo artist} \\*
   &
   &
   &
   &
   \\ \hline
\multicolumn{2}{l}{\textbf{P05} (30 - 40, M)} &
  \multirow[c]{2}{=}{{\color{ACMDarkBlue} Screenwriter}} &
  \multirow{2}{=}{Improv School / Acting} &
  \multirow{2}{*}{15 - 20} &
  \multirow{2}{*}{10 - 15} &
  \multirow[c]{2}{=}{Actor} \\*
   &
   &
   &
   &
   \\ \hline
\multicolumn{2}{l}{\textbf{P06} (30 - 40, M)} &
  \multirow[c]{2}{=}{{\color{ACMDarkBlue} Photographer}} &
  \multirow[c]{2}{=}{Health Admin Degree} &
  \multirow{2}{*}{10 - 15} &
  \multirow{2}{*}{5 - 10} &
  \multirow[c]{2}{=}{N/A} \\*
   &
   &
   &
   &
    \\ \hline
\multicolumn{2}{l}{\textbf{P07} (18 - 30, M)} &
  \multirow[c]{1}{=}{{\color[HTML]{980000} Lamps}} &
  \multirow[c]{1}{=}{Medicine} &
  \multirow{1}{*}{1- 5} &
  \multirow{1}{*}{1 - 5} &
  \multirow[c]{1}{=}{Works at university} \\*
   &
   &
   &
   &
   \\ \hline
\multicolumn{2}{l}{\textbf{P08} (40 - 50, F)} &
  \multirow[c]{2}{=}{{\color{ACMDarkBlue} Experience artist}} &
  \multirow[c]{2}{=}{Architect / artist curator} &
  \multirow{2}{*}{25 - 30} &
  \multirow{2}{*}{0} &
  \multirow[c]{2}{=}{Real estate renovation} \\*
   &
   &
   &
   &
   \\ \hline
\multicolumn{2}{l}{\textbf{P09} (30 - 40, F)} &
  \multirow[c]{2}{=}{{\color{ACMDarkBlue} Fabrication \& Construction}} &
  \multirow[c]{2}{=}{Textiles, marketing} &
  \multirow{2}{*}{10 - 15} &
  \multirow{2}{*}{1 - 5} &
  \multirow[c]{2}{=}{N/A} \\*
   &
   &
   &
   &
    \\ \hline
\multicolumn{2}{l}{\textbf{P11} (20 - 30, F)} &
  \multirow[c]{2}{=}{{\color[HTML]{980000} Greeting cards}} &
  \multirow[c]{2}{=}{Ran a few small businesses} &
  \multirow{2}{*}{1 - 5} &
  \multirow{2}{*}{5 - 10} &
  \multirow[c]{2}{=}{Tech start-up}\\*
   &
   &
   &
   &
   \\ \hline
\multicolumn{2}{l}{\textbf{P12} (N/A, M)} &
  \multirow[c]{2}{=}{{\color{ACMDarkBlue} Photography}} &
  \multirow[c]{2}{=}{Fashion and photography} &
  \multirow{2}{*}{N/A} &
  \multirow{2}{*}{N/A} &
  \multirow[c]{2}{=}{N/A} \\*
   &
   &
   &
   &
   \\ \hline
\multicolumn{2}{l}{\textbf{P13}, (50 - 60, F)} &
  \multirow[c]{2}{=}{{\color[HTML]{980000} Custom gift baskets}} &
  \multirow[c]{2}{=}{Urban and business development} &
  \multirow{2}{*}{1 - 5} &
  \multirow{2}{*}{1 - 5} &
  \multirow[c]{2}{=}{N/A} \\*
   &
   &
   &
   &
   \\ \hline
\multicolumn{2}{l}{\textbf{P14} (40-50, M)} &
  \multirow[c]{2}{=}{{\color{ACMDarkBlue} Art experiences}} &
  \multirow[c]{2}{=}{PhD electronic and electrical engineering} &
  \multirow{2}{*}{25 - 30} &
  \multirow{2}{*}{15 - 20} &
  \multirow[c]{2}{=}{Prof of digital media} \\*
   &
   &
   &
   &
    \\ \hline
\multicolumn{2}{l}{\textbf{P15} (N/A, M)} &
  \multirow[c]{2}{=}{{\color[HTML]{980000} Lamps}} &
  \multirow[c]{2}{=}{Masters in industrial design} &
  \multirow{2}{*}{N/A} &
  \multirow{2}{*}{1 - 5} &
  \multirow[c]{2}{=}{N/A} \\*
   &
   &
   &
   &
   \\ \hline
\multicolumn{2}{l}{\textbf{P16} (30 - 40, M)} &
  \multirow[c]{2}{=}{{\color[HTML]{980000} Clothing}} &
  \multirow[c]{2}{=}{Human-computer interaction design} &
  \multirow{2}{*}{5 - 10} &
  \multirow{2}{*}{5 - 10} &
  \multirow[c]{2}{=}{N/A} \\*
   &
   &
   &
   &
   \\ \hline
\multicolumn{2}{l}{\textbf{P18} (20 - 30, F)} &
  \multirow[c]{2}{=}{{\color[HTML]{980000} Stickers}} &
  \multirow[c]{2}{=}{MBA, BFA Jewelry} &
  \multirow{2}{*}{10 - 15} &
  \multirow{2}{*}{5 - 10} &
  \multirow[c]{2}{=}{Semiconductor company} \\*
   &
   &
   &
   &
   \\ \hline
\multicolumn{2}{l}{\textbf{P19} (30 - 40, M)} &
  \multirow[c]{2}{=}{{\color[HTML]{980000} Featherwing}} &
  \multirow[c]{2}{=}{Photojournalism, software engineering} &
  \multirow{2}{*}{5 - 10} &
  \multirow{2}{*}{1 - 5} &
  \multirow[c]{2}{=}{Lab Manager} \\*
   &
   &
   &
   &
   \\ \hline
\multicolumn{2}{l}{\textbf{P20} (20 - 30, F/NB)} &
  \multirow[c]{2}{=}{{\color[HTML]{980000} Clothing}} &
  \multirow[c]{2}{=}{Fashion designer} &
  \multirow{2}{*}{10 - 15} &
  \multirow{2}{*}{1 - 5} &
  \multirow[c]{2}{=}{N/A} \\*
   &
   &
   &
   &
   \\ \hline
\multicolumn{2}{l}{\textbf{P21} (30 - 40, F)} &
  \multirow[c]{2}{=}{{\color[HTML]{980000} NFTs/ Painting}} &
  \multirow[c]{2}{=}{Product Manager} &
  \multirow{2}{*}{5 - 10} &
  \multirow{2}{*}{1 - 5} &
  \multirow[c]{2}{=}{N/A} \\*
   &
   &
   &
   &
   \\ \hline
\multicolumn{2}{l}{\textbf{P22} (30 - 40, F)} &
  \multirow[c]{2}{=}{{\color[HTML]{980000} Pottery}} &
  \multirow[c]{2}{=}{Art Education} &
  \multirow{2}{*}{15 - 20} &
  \multirow{2}{*}{5 - 10} &
  \multirow[c]{2}{=}{Teaching at summer camp} \\*
   &
   &
   &
   &
    \\ \hline
\multicolumn{2}{l}{\textbf{P23} (60 - 70, F} &
  \multirow[c]{2}{=}{{\color[HTML]{980000} Glass}} &
  \multirow[c]{2}{=}{Graphic Design} &
  \multirow{2}{*}{10 - 15} &
  \multirow{2}{*}{10 - 15} &
  \multirow[c]{2}{=}{Sells vintage items on Etsy} \\*
   &
   &
   &
   &
    \\ \hline
\multicolumn{2}{l}{\textbf{P24} (40 - 50, F)} &
  \multirow[c]{2}{=}{{\color[HTML]{980000} Hair accessories}} &
  \multirow[c]{2}{=}{Made hair accessories for daughter} &
  \multirow{2}{*}{5 - 10} &
  \multirow{2}{*}{5 - 10} &
  \multirow[c]{2}{=}{N/A} \\*
   &
   &
   &
   &
   \\ \hline
\multicolumn{2}{l}{\textbf{P25} (30 - 40, F)} &
  \multirow[c]{2}{=}{{\color[HTML]{980000} Gem cutter}} &
  \multirow[c]{2}{=}{Biochemist/protein scientist} &
  \multirow{2}{*}{5 - 10} &
  \multirow{2}{*}{1 - 5} &
  \multirow[c]{2}{=}{Biochemist in Food Tech} \\*
   &
   &
   &
   &
  \\ \hline
\multicolumn{2}{l}{\textbf{P26} (70 - 80, F)} &
  \multirow[c]{2}{=}{{\color[HTML]{980000} Sustainable accessories}} &
  \multirow[c]{2}{=}{Fashion} &
  \multirow{2}{*}{50 - 55} &
  \multirow{2}{*}{50 - 55} &
  \multirow[c]{2}{=}{Floral installation} \\*
   &
   &
   &
   &
   \\ \hline
\multicolumn{2}{l}{\textbf{P27} (50 - 60, F)} &
  \multirow[c]{2}{=}{{\color[HTML]{980000} Hair accessories}} &
  \multirow[c]{2}{=}{Economics degree} &
  \multirow{2}{*}{15 - 20} &
  \multirow{2}{*}{15 - 20} &
  \multirow[c]{2}{=}{Cleans horses} \\*
   &
   &
   &
   &
    \\ \hline
\multicolumn{2}{l}{\textbf{P28} (60 - 70, M)} &
  \multirow[c]{2}{=}{{\color[HTML]{980000} Wire sculptures}} &
  \multirow[c]{2}{=}{College - no focus} &
  \multirow{2}{*}{15 - 20} &
  \multirow{2}{*}{1 - 5} &
  \multirow[c]{2}{=}{Pool cleaning business} \\*
   &
   &
   &
   &
  \\ \hline
\multicolumn{2}{l}{\textbf{P29}, (18 - 30, M)} &
  \multirow[c]{2}{=}{{\color[HTML]{980000} Tennis racket guitars}} &
  \multirow[c]{3}{=}{Musician and Creative Technologist} &
  \multirow{3}{*}{5 - 10} &
  \multirow{3}{*}{1 - 5} &
  \multirow[c]{3}{=}{Business consultancy} \\*
   &
   &
   &
   &
   \\
    &
   &
   &
   &
   &
   &
   \\\hline
   \multicolumn{2}{l}{\textbf{I01} (30 - 40, F} &
  \multirow[c]{2}{=}{{\color{ACMDarkBlue} Makerspace Co-founder}} &
  \multirow[c]{2}{=}{Learning sciences, HCI researcher} &
  \multirow{2}{*}{10 - 15} &
  \multirow{2}{*}{10 - 15} &
  \multirow[c]{2}{=}{N/A} \\*
   &
   &
   &
   &
   \\\hline
     \multicolumn{2}{l}{\textbf{I02} (50 - 60, M} &
  \multirow[c]{2}{=}{{\color{ACMDarkBlue} Craft workshop instructor}} &
  \multirow[c]{2}{=}{Information Technology} &
  \multirow{2}{*}{N/A} &
  \multirow{2}{*}{5 - 10} &
  \multirow[c]{2}{=}{N/A} \\*
   &
   &
   &
   &
   \\\hline
    \multicolumn{2}{l}{\textbf{I03} (40 - 50, M} &
  \multirow[c]{2}{=}{{\color{ACMDarkBlue} Business services}} &
  \multirow[c]{2}{=}{Bookkeeping} &
  \multirow{2}{*}{N/A} &
  \multirow{2}{*}{10 - 15} &
  \multirow[c]{2}{=}{N/A} \\*
   &
   &
   &
   &
   \\\hline
     \multicolumn{2}{l}{\textbf{I04} (30 - 40, M} &
  \multirow[c]{2}{=}{{\color{ACMDarkBlue} Entrepreneurial hub co-founder}} &
  \multirow[c]{2}{=}{Business school} &
  \multirow{2}{*}{N/A} &
  \multirow{2}{*}{5 - 10} &
  \multirow[c]{2}{=}{N/A} \\*
   &
   &
   &
   &
   \\\hline
       \multicolumn{2}{l}{\textbf{I05} (20 - 30, M} &
  \multirow[c]{2}{=}{{\color{ACMDarkBlue} Community manager}} &
  \multirow[c]{2}{=}{Facilities and operations} &
  \multirow{2}{*}{N/A} &
  \multirow{2}{*}{1 - 5} &
  \multirow[c]{2}{=}{Nutritional spice business} \\*
   &
   &
   &
   &
   \\\hline

       \multicolumn{2}{l}{\textbf{I06} (40 - 50, F} &
  \multirow[c]{2}{=}{{\color{ACMDarkBlue} Incubator director}} &
  \multirow[c]{2}{=}{Creative writing} &
  \multirow{2}{*}{N/A} &
  \multirow{2}{*}{1 - 5} &
  \multirow[c]{2}{=}{Creative writing coach} \\*
   &
   &
   &
   &
   \\\hline

       \multicolumn{2}{l}{\textbf{I07} (50 - 60, F} &
  \multirow[c]{2}{=}{{\color{ACMDarkBlue} Art curator director}} &
  \multirow[c]{2}{=}{Curated galleries} &
  \multirow{2}{*}{N/A} &
  \multirow{2}{*}{1 - 5} &
  \multirow[c]{2}{=}{N/a} \\*
   &
   &
   &
   &
   \\\hline
\\

\end{longtable}
\end{footnotesize}

%% file: 03_results.tex
\section{Results}
\label{sec:results}

{In this results section, we divvy up the findings from the data analysis as they relate to the guiding research questions from} \autoref{sec:intro}.{The first two research questions, which focus on the transition from maker to maker entrepreneur, and the business challenges unique to maker entrepreneurs, are important because they help us understand the \textit{work context} of making for a living, as opposed to for a hobby, {and provide motivation for technology design considerations.}
The third question, which addresses the role of technology in maker entrepreneurship, is then critical to understanding what CSCW {technology} can do for this community.}

\begin{figure}[ht]
 \centering
\includegraphics[width=0.9\linewidth]{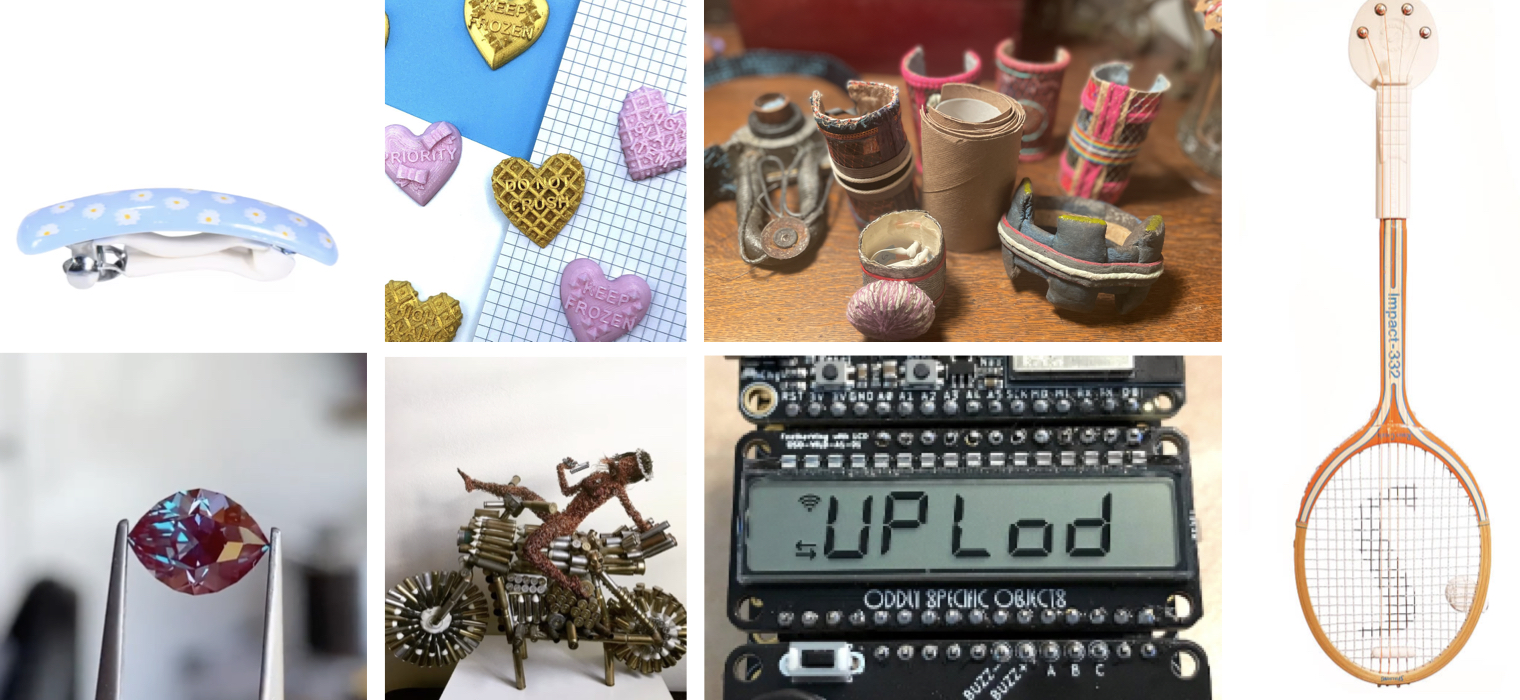}
\caption{Variety of maker products: \textit{Top left:} P27's hair accessory; \textit{Top middle:} P18's 3D printed stickers; \\ \textit{Top right:} P26's cardboard cuffs; \textit{Bottom left:} P25's Gemstone; \textit{Bottom middle:} P28's wire sculptures; \textit{Bottom right:} P19's Featherwing; \textit{Rightmost:} P29's guitar tennis rackets}
 \label{fig:examples}
\end{figure}

\input{03a_results_RQ1}
\input{03b_results_RQ2}

\input{03c_results_RQ3}

%% file: 03a_results_RQ1.tex
\subsection{What is involved in the transition from maker to maker entrepreneur? (RQ1)}

 Many maker entrepreneurs we interviewed were ``accidental entrepreneurs'' \cite{aldrich1999accidental} in that they did not set out to become business leaders but evolved into them after initial commercial interest in their work as a craftsperson or hobbyist. 
 Table \ref{tab:participants} shows that {15 of our 26 interviewees have art or crafting experience prior to starting their business.}
 Some of these entrepreneurs had full-time jobs but whose craft drew a growing demand from friends and family. For example, P12, a fashion photographer, talked about how they started their work by taking photographs of their family with no initial interest in entrepreneurship: \pquote{I was like, I want to do a fashion course in arts. So I started by taking a photo of my parents and family, and it gradually grew...I didn't plan to make it a business.}
 
Our interviewees spoke about how their transition into business often aimed to support their primary goal of being a maker. For example, P25, a gem-cutter, began to sell their gems so that they could cut more and thus continue with their hobby in a monetarily sustainable way.
P08, an artist, similarly explained: \pquote{I think the main challenge is to find a way to combine those two [identities] so I will be making money out of the things I like to do...}
{However, it was often challenging for makers to figure out how to make money doing what they love. Overall, the makers we spoke with, who often did not plan to become entrepreneurs, discussed how they needed to reconcile their maker and entrepreneurial identities, while also learning to manage multiple business roles they did not have training for.}

\subsubsection{Reconciling maker and entrepreneur identities}
\label{sec:recon}

A recurring theme in our interviews was the tension maker entrepreneurs felt between making and entrepreneurship {and the values that they attempted to prioritize.}
For some, entrepreneurship was a means to the end of making---{5 out of 20 makers tended to explicitly look for ways to prioritize or give more time to their craft, rather than maximize financial reward.} 
P22, a potter, said, \pquote{I wanted to sell my art first and I did it. I wanted to get the artist's residency and I got it. I wanted to put my pieces in a gallery, which I got it. But my goal now is to put my pieces in a museum in the future. That is my 10-year goal.}
{For P22, the financial gain of selling their work allows them to work for their higher artistic goals.}

Other interviewees were quick to proclaim {other} priorities they put ahead of money.
For instance, one of our consultants, who runs a business incubator, I06, said: {I struggle with that phrase, [entrepreneurship]. I think I'm an anti-capitalist in my heart of hearts. I don't have any interest in being a millionaire or in helping other people to become millionaires... Like, I think that there are a lot of reasons that people would not want to scale a business or make millions with a business, and that doesn't mean that they shouldn't get support for it.}
{In this, I06 recognizes that many maker entrepreneurs may want to run a stable business, but may not have goals to grow large, suggesting maker enterpeneurs with different values may need sufficient, yet a different kind of support.}

While turning a profit and running a sustainable business need not be in opposition to creative maker work, makers faced challenges combining the maker and business identity, psychologically complicating the enjoyment of creative activity. 
P07, a lamp maker, stated: \pquote{I'm trying to figure out, like, do I make products to sell to people or do I just make things for my own kind of happiness and fulfillment?} 
For some, such as P11, a stationery maker, this tension ultimately led to their reverting to making as a hobby rather than a business venture: 
\begin{quotebox}
    So I've kind of since turned that into more of a hobby...rather than business...That kind of took away the joy of what I was trying to do with it, and it just like became too stressful, too much pressure---too much expense upfront, with very little knowing that I could actually make that money back.
\end{quotebox}
These quotes suggest the tension makers may feel when figuring out how to balance their creative goals and enjoyment with the need to run a sustainable business.

\subsubsection{Managing time and multiple roles}
\label{timeManagement}
{Part of what contributed to our interviewees needing to reconcile their identities as makers and entrepreneurs included the various roles they needed to take, with increased time devoted to business management}. 
Our interviewees discussed balancing multiple responsibilities beyond their creative work: they needed to manage and organize money, market their products, manage personnel, and navigate physical logistics. This called for appropriate time management and the ability to prioritize between creating and selling. 

However, making creative goods takes a considerable amount of time, with some participants describing how they often could not estimate at the outset of a project. 
P21, the abstract artist, explained, \pquote{I never know how much time it will take...you re-work a lot...you never know when it is finished.} 
P28, a wire sculptor, also took time to look at his sculptures in the light in his living room for hours: \pquote{Sometimes I come, I turn the light on, and I'll look at it, for about a week...[I] look at it...[and give it ] a few touches..it makes a difference in my eyes.}
P07, a lamp-maker, mentioned, \pquote{``having enough time to make things, I often like hyper-focus on something and I kind of ignore other things but it is challenging to.}
By default, our makers often focused their goals, attention, and time on making, suggesting that this usually meant the entrepreneurial aspects of their livelihood were made more precarious.

{The challenge of balancing making money with their craft meant that makers considered different configurations for their overall work.}
Many of our interviewees had other jobs in addition to their maker business, but stated that the allocation of time and energy for both jobs was a concern.
Some interviewees wished they could work on their business full-time (P07, P11, P27) and usually focused on being financially independent as a goal. 
Hair accessory maker P27 stated, \pquote{Success for me ultimately will be able to do this full time.} 
This being said, other makers, like P25, a gemstone cutter, preferred their making as a side job because it created a balance with a full-time job in another {non-making industry, while still allowing them to financially support their craft}.
{Overall, the balancing of creative and business tasks toward varying personal goals suggests that makers may need more personalized support in working on their business.}

\subsubsection{Grappling with business basics}

\label{sec:AccountingAndTaxes}
Our interviewees' difficulties reconciling their values and priorities in entrepreneurship, and their "accidental" entrance into entrepreneurship, also seemed to affect their business operations. By and large, the maker entrepreneurs in our study did not keep consistent records of their receipts, production costs, and taxes at the early stages of their business due to a ``side gig mindset.'' A recurring theme in our results was maker entrepreneurs managing the business side of the work in an ad-hoc manner. 
For example, P07 tracked payments from customers through multiple platforms which he manually updated in an application not originally intended to support business transactions: \pquote{they just Venmo me... and then I keep their names in a [iPhone] Notes app.} 
The entrepreneurial hub co-founder, I04, with finance expertise, {corroborated} that accounting was a particular shortcoming with many creative entrepreneurs: 
\begin{quotebox}
    Nobody thinks about accounts receivable, taxes, and all of these things that really matter... it's beneficial to think about those things starting out, instead of only thinking about the logo...
\end{quotebox}

A few of the maker entrepreneurs we interviewed seemed to have learned through the mistakes they made in the early phase of their development. P02 explained that he wished he had a budget before starting anything because he spent too much money on personal items while starting his business: \pquote{You are spending way too much money on vinyl, or on eating out. Where's all this money going? And so, you have to be intentional... [I learned to] have a budget before you start anything.}
P06, a photographer, talked about their naivete regarding payment: 

\begin{quotebox}
    When I first started kind of just making money, it was just strictly cash. I didn't know how to write invoices. It was just, pull up to the shoot with the cash ready... I learned over time that the proper way to do business is to learn how to write invoices.
\end{quotebox}

{Overall, the maker entrepreneurs we spoke with that had developed better business processes often did so as they carried out their business activities while figuring out how to be an entrepreneur.} 

%% file: 03b_results_RQ2.tex
\subsection{What are business challenges unique to maker entrepreneurs? (RQ2)}

{While all entrepreneurs can experience challenges running their business, we aimed to learn what challenges maker entrepreneurs in particular might have.} 
More specifically, we noted makers had difficulty finding economically sustainable ways to make their goods, connect to their clients, price their wares, and connect to communities.
{These findings provide context from some of the technology challenges we discuss in Section \ref{sec:findings-tech}, especially around the different Internet marketplaces (i.e., eCommerce platforms) our interviewees were often challenged by.}

\subsubsection{Challenges of small-scale physical production}
Many of our interviewees spoke about how their handmade craft production led to unpredictability around the time it took to make items.
For some entrepreneurs, reliance on specific, expensive physical equipment kept them tied to a particular studio location and working around that studio's schedule. 
For instance, P22, a potter, shared: \pquote{If you don't have your own kiln, you have to go somewhere to ask for firing. But they have their own schedule, so sometimes it takes a really long time.} 
Their process of making physical goods also resulted in breakage and damage from time to time. 
P22 continues: \pquote{So when the plate is not dry enough, it blows up. And when the fire doesn't go up enough, the color is weird. And if the pieces touch, it melts together.} 
This risk, combined with relying on another studio's kiln firing schedule, slowed the process and put pressure on the maker entrepreneur to not make mistakes that were costly in terms of both time and money.

However, small-scale physical production gave the maker entrepreneurs we spoke to a sense of fulfillment and a satisfying connection to the made artifact, {again speaking to the values they may have over pure financial growth.}
P25, a gem cutter, describes that tangible creations give her satisfaction outside of her slower-paced, full-time job as a scientist, where results took time:
\begin{quotebox}
You know, gem cutting, I can sit down in an evening, finish a gem, hand it off to a satisfied customer, you know, they come back in a few months and show me what they've made and it's very fulfilling to see from the beginning to the end and you see you have a tangible impact on someone's life.
\end{quotebox}
{Thus, changing production processes may not be a desired option for makers as they grow their business.}

\subsubsection{Connecting to clients}

 Many of the makers we interviewed found it challenging to find the right audience of customers who would appreciate and purchase their products.
 P22, a potter and an immigrant from Japan, had limited resources when it came to finding places to sell her work. {Leveraging resources at hand,} she did market research through Facebook reviews: \pquote{I didn't have any information about art markets. I just Googled and tried to see all of the reviews... I post on a Japanese Facebook group that shares (large US city) information in Japanese.} 
Others relied on their products being photogenic {and} leveraged social media and the press as a way to reach customers without the need for targeted art or craft venues.
{However,} while social media exposure could be positive for their visibility, it did not always translate into increased sales. 
For example, P07 questions, \pquote{When does exposure become money? That's really just what I need...I've seen a post I made go viral and get like a hundred thousand views and it was exciting but also, the internet has moved on.} 

However, the right client {who finds the makers work} could be transformative for the maker's business. 
P29, a maker who turned vintage tennis rackets into guitars, described one such case:
\begin{quotebox}
    I'm hoping that I can build up a following; where people go, why the hell did you make this? Or, holy shit, I want one! ...you're not trying to sell to everyone immediately. ...I'm gonna give one to my artist friend who's gonna make a bunch of content with it. So he's the guy that's gonna make a song out of it. He's the guy. He has a big following.
\end{quotebox}

Other maker entrepreneurs ``shopped'' for clients who could afford their products to support a sustainable business model. P12, for example, described trying to attract wealthier clients: 
\begin{quotebox}
    Our goal is to be able to attract [wedding] clients that are willing to pay us more. We know that those clients are also willing to invest more into the [photography] open-ended planning process and that will influence the quality of our work.
\end{quotebox} 
Similarly, the gift basket maker, P13, customized her baskets based on how much customers could pay.
{Ultimately, these varying strategies suggest that maker entrepreneurs took different strategies in finding customers over just connecting with people admiring their work.}

\subsubsection{Pricing their goods}
\label{sec:determining}
{In addition to finding paying customers,} entrepreneurs also had a variety of strategies for pricing their goods; they looked at the perceived value in the market, the size of the piece, client spending power, and the cost of materials and sourcing location. 
 
The makers we spoke with, however, had unique challenges because craft goods are somewhere between the spectrum between commodity goods, which are priced based on market value, and art, which is priced based on the symbolic value \cite{khaire_culture_2017}---far more elusive to pin down. 

One of our interviewees, P27, the hair clip designer, described the straightforward process she used to price her hairclips: she benchmarked pricing patterns in the industry through her experience shopping in the USA. She sourced her products (for re-design) from Alibaba and found that products that looked expensive were cheap. She therefore made things where the cost of production would be lower than the perceived market value.

However, pricing was often far more complicated for our other interviewees, often again due to their own personal values. 
P13 described not wanting to rip people off by significantly increasing the price of their product, but stated that they did need to mark it up enough so there was a reasonable profit. 
{Maker entrepreneurs also described other challenges} in pricing their work, such as pricing pressures from commodity markets, even those intended for maker entrepreneurs. 
P23, a glassmaker, described that Etsy, in response to Amazon offering free shipping, encouraged its sellers to offer free shipping and move the cost into the price of the good.
While consumers liked free shipping, they described how this could eat into the maker's profit because they did not want to increase the price more to cover the shipping and risk losing sales.

We saw some indications that successful maker entrepreneurs we spoke with were able to reason about and capitalize on the non-monetary value that customers derive from maker goods.
 The painter P02 discussed the values associated with nostalgia: 
\begin{quotebox}
    You know, they'll play it on a cassette, and they'll be nostalgic, and maybe they'll buy more cassettes. So it's like, people buy this for \$15 right now [and] they keep this. This is like a low-priced item. They keep it for 10 years. And as I'm working and doing my thing, and it's like, oh, shoot, like you had his first mixtape.
\end{quotebox} 
P02 understood that items he made could accumulate value over time because of nostalgia.
 
 P09 also described the way that consumers purchased items as an expression of their values: 
 \begin{quotebox}
     But like Tom's\footnotemark Giving away shoes if you buy shoes like really makes people feel good...Impacting somebody instead of, you know, Amazon or Apple. And then I think, yeah, we express ourselves through our consumption.
 \end{quotebox}
 \footnotetext{Tom's Shoes is a company that donates a pair of shoes for each pair sold.}

 {Building on this idea of thinking about pricing with regard to how customers express themselves,} I02, who ran hands-on workshops for various maker activities such as woodworking, knife-making, and jewelry, described his value to his customers as selling them ``bragging rights'': 
 \begin{quotebox}
     They get to experience what it's like to weld even though they're a marketing professional who dresses nicely and goes to respectable places. They come here and get to make a lot of noise and sparks... hopefully their friends notice [the thing they made], and they get to brag on themselves and say, I made that.
 \end{quotebox}
Through this experience, customers got to tell their friends about their new skills as well as develop a new sensibility about the products they bought.
 {Based on these comments from our interviewees, there is an opportunity for maker entrepreneurs to develop the sophistication to reason about the non-monetary value that customers derive from their goods. With such sophistication, maker entrepreneurs can better design products that engage with these values, such as the nostalgia, bragging rights, and self-expression mentioned above.}

\subsubsection{Building a community}
For maker entrepreneurs, community and relationships were interwoven across their personal identity and their business brand. Building a community of friends and followers {who admire their work} on social media doubles as a potential customer base and enables them to create close relationships with customers, suppliers, and business partnerships.

Our interviewees recognized the importance of community and discussed how they created community (P02, P04, P06, P09, P11, P12, P20, I04). 

For instance, P20 described their relationship with their manufacturer: \pquote{We kind of have a rapport or a relationship that feels...like a closer friendship}. 
For P20, what started as a strictly business relationship had evolved into a friendship, providing several benefits, such as ensuring the authenticity of the relationship, aesthetic alignment, and business commitment. 
For P25, a gem cutter, building a relationship with a potential gem dealer was critical since her small operating scale usually made it hard for her to be taken seriously: \pquote{[I was trying to] build up those dealer relationships, saying that, I can be a consistent customer. I have consistent sales, like take me seriously.} 
P02 reflected on his experience building community, which he did with an eye towards evolving those communal relations into business transactions: \pquote{I am at the [gallery], shaking people's hands, just showing people that I do dope stuff and [they become] a part of my community...}
P02 went on to discuss how this community-building approach was critical to ``solidify the bag,'' or ensure that interested onlookers converged to paying customers.

Participants reflected on these kinds of relationships which required balancing genuine relationships that were personally fulfilling while also ensuring business success.
One curator, I07, we spoke with described the ideal: \pquote{I wanted to promote artists and wanted to have a great...connection and relationship with artist[s]} {suggesting how positive connections could lead to more business through brokered promotion.} 

However, for some entrepreneurs, relationships could interfere with a more objective assessment of their products. 
As P11 described: \pquote{I just wanted to test [my idea] out without having my name seeded with it so that people wouldn't feel like, `I'll do this to support [P11's name].'} 
Here, P11 described how they struggled to garner honest feedback {from their community} that was not cushioned with friendly support, making it harder for them to test if their idea had true market viability. 
Such signals {of customer desire} were critical for maker entrepreneurs, who, due to their small-scale, had limited time to dedicate to research and development.
{Our findings suggest that systems to support maker entrepreneurs through community should be designed to assist creating connections that support business, beyond simply making connections between people and the maker.}

%% file: 03c_results_RQ3.tex
\subsection{What role can technology play in supporting maker entrepreneur business activities (RQ3)} 
\label{sec:findings-tech}
{With an understanding of the context in which maker entrepreneurs work, we then sought to learn about how technology supports them.}
Many digital tools exist to support general business processes such as learning to start a business, acquiring customers, marketing, selling, and managing finances and business operations \cite{hisrich2017entrepreneurship,openstax_entrepreneurship_2020}.
The Maker entrepreneurs in our sample were not strangers to such tools and used a variety of existing digital tools to support their businesses.  
However, our interviewees often pointed out shortcomings of many popular tools and platforms and described how they formed their own solutions best for their business at its current state.
Interestingly, many recognized and desired more support from digital tools, but still chose not to use the ones available.
These findings highlight technology opportunities for business tools to specifically support maker entrepreneurs.

\subsubsection{{Learning entrepreneurial and business skills}}

{Like many other entrepreneurs and especially other arts-based creatives} \cite{makridis2023narrowing}, our maker entrepreneurs often had no formal training or education related to business, even though many had college degrees.
However, instead of devoting time to complete an overview business course or degree like an MBA, our maker entrepreneurs often sought out information as needed from non-academic, accessible resources like online videos, the library, the internet, and social network to learn how to run their business.
P06 stated when describing how they learned how to run their business, \pquote{I went to YouTube University.} 
P27 went to the library to learn how to file a patent application and also received feedback from a friend about patent office common practices.
Even those who had taken formal entrepreneurship courses in university still maintained mentors to help them as they built their business, such as P15 who described his ongoing mentorship from their entrepreneurship professor: \pquote{He has someone who's been around the business of design for many many years...he can see around corners that I don't even know are there.}
{Overall, the maker entrepreneurs assembled their own resources and advisors to suit their specific needs.} 

{One particularly salient example of how maker entrepreneurs created their own learning resources also intersects with how digital tools are used to create online subcommunities.}
P25 described how they engaged with and contributed to a Reddit group for gem cutters, which focused on advice for running their unique businesses given the nature of producing cut gems, sourcing materials from global suppliers and making international payments, and growing their customer base using social media platforms.
In this specific case, P25's community had shaped their subreddit with explicit rules for posting educational material rather than selling gemstones. 
This allowed the subreddit to become a place {focused on learning and personal growth,} where gem cutters could share their craft, critique each other's work, and advise on running a gem-cutting business with strategies for working with suppliers, formally registering a business, pricing, taxes, and getting insurance. P25 describes, \pquote{Over time, I pivoted a little bit away from the solo Instagram model to more investment in the subreddit for the community aspect of it.} This example highlights how some maker entrepreneurs may be able to create online communities to support each other's learning, but also raises questions as to how many other maker entrepreneurs creating different goods do not have access to such resources and where online technologies and platforms could better support these needs.
Furthermore, the collection of resources and communities that maker entrepreneurs use show that they build their business education just-in-time to suit their needs at whatever stage of their entrepreneurial journey they are on.

\subsubsection{{Finding customers}}

To promote and market their goods, the makers we spoke with primarily used social media to show their work visually and have a direct communication channel with their potential customers.
A majority of makers used Instagram to post photos of their completed work.
These posts were both a way to advertise their work, with links to purchase through an eCommerce platform or via direct messaging.
For makers who produced individual items, Instagram also functioned as a digital gallery space where many people could enjoy the work even though only one customer would be able to purchase it.
{This allows makers to create a fan base who support their work in general, increasing the value of all their work and building more engagement which is helpful on Instagram for promoting posts beyond the maker's followers.}

{Another interesting and maker-specific behavior that many described was to show photos and videos of work in progress while working in their shop.}
{Makers described how these kinds of posts helped them} tell their story and present their identity as a maker and solo entrepreneur.
One participant, P19, described using Twitter (X) to document their work in progress and build excitement about their upcoming electronics board.
Sharing work in progress, with all the trials and successes, had the potential to raise the value of the final product since potential consumers saw the effort that went into it and were a part of the story, similar to how crowdfunding sites can help people build a following around their products \cite{gerber2013crowdfunding}.

{However, while makers were aware of the algorithms that could promote or bury their posts and regulate communication with their fans and customers, they were not always keen to engage with specific posting and promotion practices.}
Some maker entrepreneurs described using platforms that were less algorithmic, so that they could have more control over their communication with customers and fans.
P25 described how they used Reddit to promote their work because the Instagram algorithm could change and reduce the number of people who see their work.
This in turn could lead to a reduction in sales that the maker had little to no control over, since they did not understand how the algorithm had changed.
They also described how the design of commenting on Instagram could limit engagement:
\begin{quotebox}
    [The Reddit] community is so much more of a draw to me than just posting on Instagram by myself... I really like being able to better interact with clients and have that platform. In Instagram comments, if you write a comment, you're just scrolling, you see like two comments, and it's really hard to follow threads.
\end{quotebox}
P25's challenges with Instagram were echoed by P07, who described Instagram as primarily being a one-way communication tool for marketing and lamenting that they felt there were no great platforms for two-way communication between maker entrepreneurs and their customers.
{For these maker entrepreneurs though, such communication was important for finding customers and marketing their work.}

\subsubsection{{Selling Products}}
{Given the {boom} on digital eCommerce platforms on the web today, and especially more {specialized} marketplaces for particular product categories, we expected many maker entrepreneurs that we spoke with to use a variety of online marketplaces.} 
These included platforms that provided online ``shops'' within a larger category of similar goods, such as Etsy (Vintage \& Handmade), Tindie (DIY Electronics \& 3D printed products), and Crowdsupply (Electronics).
Online maker goods platforms such as Etsy and Tindie can provide makers exposure to customers looking for craft products in specific categories.
As described by P19, who sold DIY electronics: \pquote{Tindie is nice because you're on a platform with other products that are maker-oriented. And there's definitely an understanding that this is not like going to Best Buy and getting something that's fully formed.}
Such platforms attempt to highlight the individual creative value of a maker rather than be a store for commodity goods while still providing a platform for customers looking for specific kinds of products.

However, despite these platforms being clearly designed for maker entrepreneurs, among our interviewees, only four were currently using platforms such as Etsy or Amazon Handmade, while 15 used a self-managed solution such as Shopify and Squarespace or direct social media messaging/in-person sales with a peer-to-peer digital payment system (i.e., Venmo / Paypal).
We found that makers had challenges with online platforms, such as dictating terms requiring free shipping on all orders or not supporting important financial functions such as adding sales tax to orders. 
P19, quoted above about the benefits of a platform like Tindie, eventually moved away to self-manage their store using Shopify purely due to business rationale (e.g., Tindie did not include tax-withholdings) even though they liked and appreciated the community around Tindie. {P19 reluctantly describes,} \pquote{{Shopify, at least handles sales tax.}} {P19 is not happy about the tradeoff, as they have to trade the artistic community of Tindie for simplified business administrative tasks.}

Much in the way that makers wanted to create products around their individual creativity, such examples also suggest that while eCommerce platforms do support many makers, our result suggest there are unmet needs around specific functions or simply managing their businesses on their own terms and prompt makers to tailor their sales solutions using a mix of technologies on their own.

\subsubsection{{Managing finances \& operations}}
While all businesses that sell physical goods need to track their finances, inventory, and communications, many of our participants found technology solutions lacking for specific aspects of their maker business.
Regarding financial tracking, we found that {participants using more formal sales tools used the financial tracking that came with them.} 
Some participants mentioned using professional tools such as Quicken\footnote{https://www.quicken.com} (P23) and Manager\footnote{https://www.manager.io} (P19) for tracking financial information, and one participant managed their own financial spreadsheet; however, these were exceptions among our interviewees.
P27, the hair-clip designer with a background in economics, was an anomaly, as she enjoyed doing taxes and keeping her receipts because it gave her control of her business. 
She recommended that all entrepreneurs should keep a shoe-box\footnote{Storing receipts in a shoe-box is a common strategy among personal finance recommendations.} with their receipts. She explained, 
\begin{quotebox}
    {Quicken makes something where you can just take a picture....It's all on your phone and [categorizes]  what kind of expense it is...But I think if you don't do it yourself, you don't know what receipts to save.} 
\end{quotebox}
{She alludes to an important point here - that an efficient tool does not necessarily create financial knowledge. Instead, financial knowledge supports the use of efficient tools.}

Even when makers did use {digital} tools to track their finances, they often did not include key financial information such as the cost of materials or time spent making an artifact. 
The lack of key financial details meant it was hard for maker entrepreneurs to understand their profit or which products might be cash cows versus loss leaders.
Many makers noted that taxes could be quite challenging to manage due to the limited financial tracking. 
Financial tracking was further complicated by the fact that many did not use separate business bank accounts, utilizing one (personal) account for all their finances.
{The complexities and limitations of many tools, along with limited business education, likely led many of the maker entrepreneurs we spoke with to simply not use more common financial management tools and instead cobble their own processed (if they had them) together.}

%% file: 04_discussion.tex
\section{Discussion}

From a CSCW perspective, our interest lies in creating information technologies that can better support {working} maker entrepreneurs in achieving their goals. Looking at the results, it is straightforward to see opportunities in broadening access to technologies currently being used by maker entrepreneurs (as discussed in findings for RQ3) by providing just-in-time learning, enabling direct relationships with customers, and managing sales and finances. However, the responses for RQ1 and RQ2 suggest more meaningful challenges that we feel CSCW researchers can also help to tackle. Here, we discuss three such areas.
{For a brief overview of these categories and implications for design, please see Table} \ref{tab:implications}.

\subsection{Just-in-time training tools can address knowledge gaps created by "accidental entrepreneurship."}

A recurring theme throughout the interviews was the way that maker entrepreneurs had to learn things on the fly or teach themselves business skills. 
This occurs in part because they come into entrepreneurship "accidentally" and sometimes as a part-time activity \cite{aldrich1999accidental}. 
As a consequence, maker entrepreneurs lack formal business training and do not know about different business structures, such as an LLC, or what they might need to operate their business, such as accounting systems or insurance. 
Instead, the maker entrepreneurs in our study learned the ropes through trial and error, just-in-time self-directed informal learning from resources like YouTube or workshops.
While this autodidacticism is admirable, the piecemeal nature of the learning can lead to gaps, and fail to prepare makers properly \cite{afroze2014women,shultz_work_2015}. 
Educational and training resources for maker entrepreneurs should {take inspiration from and} maintain the key benefits of "YouTube University"--- bite-size modular lessons to remedy issues "just-in-time"--- but do more to point out to maker entrepreneurs what they need to learn. {Community networks of peers, for example in }\citet{kotturi2024peerdea}'s Peerdea or \citet{wenger1999communities}'s communities of practice, might help to provide information where and when it is needed. Additional curation or modular curricular development (for example, the just-in-time entrepreneurial training suggested by \citet{sullivan2000entrepreneurial}) {could help make sure that the resulting knowledge of business skills is more comprehensive.}

Moreover, our findings indicate entrepreneurial curriculum and training for makers should include a greater emphasis on the management of operations and finances.
Many of our interviewees had less-than-ideal accounting systems and often did not know what was important for accounting until they were met with a challenge, such as having to prepare their taxes and becoming overwhelmed by piecing old transactions together.
{Maker tools could potentially support teaching important concepts at key moments, like tracking expenses and sales for tax purposes after a maker's first sale on a platform.}

Beyond accounting, maker entrepreneurs may need to pursue marketing and client development differently from other kinds of products \cite{shultz_work_2015}, aiming to find specific customers who will buy and promote a product to help increase its symbolic value \cite{khaire_culture_2017}.
Maker entrepreneurship education may focus more on helping teach makers how to tell their story, connect directly with customers to foster a fan base and learn to find customers who may be able to pay more and better promote and add value to their products.

\subsection{But why does entrepreneurship have to be accidental?}

{One of the challenges to knowledge acquisition for our interviewees, however, was not the absence of materials or tools, but rather difficulty recognizing that they needed them. This is due to that "accidental" nature of the transition. To this end, we ask whether that first step towards identity formation is not the more critically needed provision.}

 \subsubsection{{Supporting early-stage maker entrepreneurs}}
{Prior to the help that maker entrepreneurs need for performing entrepreneurship, however, is the help that these maker entrepreneurs need to recognize that they are becoming entrepreneurs. In RQ1, we looked at the transition from maker to maker entrepreneur and found that the makers we interviewed lay upon a wide spectrum in their business journey. As } \citet{dillahunt2018entrepreneurship} {described, current technologies disproportionately benefit individuals with past entrepreneurship experience. Most of our maker entrepreneurs were first-time entrepreneurs and had similar experiences and challenges with business technologies, leading many to simply to use them. }

{Business support technologies should assist those without prior entrepreneurship experience. For example, these technologies can use language that creatives or people with low entrepreneurship experience can understand. Other CSCW researchers} \citet{kotturi2024peerdea} {have focused on the learning feature of a platform for creative entrepreneurs, which implements scaffolding. We suggest similar approaches, with educational technology design taking into account the person's entrepreneurship experience and the stage of entrepreneurship they are in.
Approachable and educationally focused business tools could help makers in forming higher self-efficacy in their business skills and thus promote forming a positive entrepreneurial identity.}

\subsubsection{ {Mediating multiple identities}}
{Having multiple roles while operating the business was challenging for people's perception of themselves} (\ref{sec:recon}).
The fact that a large portion of these maker entrepreneurs had other jobs complicates their identity formation.
Maker entrepreneurs also faced frequent challenges in juggling multiple responsibilities at once. 
For example, they tracked finances, did physical labor to make craft objects, and marketed their product to potential customers. 
Such activities may be considered separate from one's creative identity, {which} \citet{glaveanu_creativity_2014} {has described as a creative practice which is perceived by the self and others. 
Tensions arise when maker entrepreneurs focus on their creative practice versus their business tasks} \cite{gotsi_managing_2010}. 
The struggle to reconcile their passion for creative work with the practical challenges of running a business led makers to question whether to pursue their craft as a hobby or a source of income. 
Business researchers \citet{gotsi_managing_2010} {have explained a way to mediate these tensions. 
They suggest an integrative framing of their identity as a `practical artist' rather than segregating creative and business identities. 
By designing technology and learning materials with this integrative framing, tools can encourage an identity which is more comfortable to maker entrepreneurs. 
For example, designers of an entrepreneurship support application could implement tabs, titled, ``practical'' and ``creative'' to acknowledge that artists have these identities but help create a positive and integrative framing around these identities.}

{Overall, support technologies should be designed in a way that acknowledges the diversity of goals that maker entrepreneurs might have.}
\citet{markman2003person} {explain that people who have a higher sense of self-efficacy tend to be more successful in their entrepreneurship endeavours.} 
Practices and tools to encourage value-elicitation, reflection, and goal-setting could help maker entrepreneurs be more mindful of what they are trying to achieve, especially when these aspirations deviate from more traditional and capitalist objectives.
This echoes findings from prior work about the challenges creatives have in developing a business identity \cite{glaveanu_creativity_2014,gotsi_managing_2010}. 
Prior work indicates that self-reflection and interaction with other creative entrepreneurs can help maker entrepreneurs reconcile the need for entrepreneurship as a support to their larger quest for creative autonomy and success while remaining authentic to themselves \cite{werthes_cultural_2018}. 
Designers of tools and training experiences should consider how creative and entrepreneurial tasks are managed and how creative and business identities are reconciled \cite{gotsi_managing_2010}, {possibly though designing interactions that promote and track self-reflections and the maker entrepreneur's identity growth.}

\begin{table} [t]
\caption{Summary of implications for design inspired by findings.}
\label{tab:implications}
 \centering
 \small 
\begin{tabular}{ | p{5em} | p{14em} | p{21em} | } 
\multicolumn{3}{|c|}
{\textbf{Implications for the design of support tools for maker entrepreneurs}}\\
 \textit{Theme} & \textit{Implications} & \textit{Examples} \\
  \textbf{{Just-in-time learning}} 
 & Designers should consider that tools for training and business support need to be granular and modular to account for the individual needs of maker entrepreneurs, and they also need to be dynamic and change along with how the business changes. 
 & {Learning systems can provide just-in-time modules as short video lessons around business concepts}  \newline \newline{Business tools could include learning modules to teach important concepts like managing taxes, at key moments like a maker's first sale.} \newline \newline{{A business support application could have an FAQ created by the community and augmented with an AI assistant.}}

 \\ 
 
 \textbf{{Identity formation}} 
 & Designers should consider that identity support is an important requirement for tools that are developed for maker entrepreneurs. Identity support tools and efforts could happen both at the individual and community level. Designers could learn from offline community support to envision how to bring this experience online.
 & Tools to support identity development might help makers track their progress as they transition to entrepreneurship, through prompting reflection about maker's self-efficacy and new business skills. \newline \newline {In a support application, external links to online communities filtered by language, business experience, and industry can support makers in connecting with others who can help them during their transition.}

 \\ 
 
 \textbf{{Tracking alternative values}}
 & Designers should consider that maker entrepreneurs are often motivated by non-monetary values such as happiness and engagement with the community. Tools for maker entrepreneurs could be developed to take their values into account or help them track their value alignment.
 & Taking inspiration from fitness trackers and journaling apps, {applications could be developed to help makers keep track of their values, such as ``creative expression'' or ``positive community engagement'', and map these alongside other business values like, ``sales'' or ``expenses.''} \newline \newline {Tools can help makers reflect on their values and let them create their own categories and metrics for tracking them.}
 \\

\end{tabular}
\end{table}

\subsection{Can accounting tools help Maker Entrepreneurs track the things they value?}

One of the recurring themes through the answers to RQ1, RQ2, and RQ3 was that makers valued many things beyond financial gain.
{Similar to findings by Annette} \cite{homepreneurship}, {many of the maker entrepreneurs we spoke with said their businesses formed to support their craft and as a money-making activity over more conventional employment. 
Building on their findings, our work highlights a set of other non-monetary values that motivate maker entrepreneurs.}
More specifically, these include entrepreneurs' values of creative expression, promoting values such as sustainability and craft, operating in ways different from mainstream capitalism, and forming relationships with customers and fans of their work.
These motivations can drive people's business decision-making {in ways that are not profit-maximizing.}
{For example, some of our participants noted producing items that will resonate well with their online community.
Such items could help build a fan base and fulfill a community and creative expression motivations for the maker, even if producing items for this purpose might not translate directly to more sales.}
{These alternative values demonstrate a difference from the generally monetary motivations of general micro-entrepreneurs while underscoring prior research on other alternative values beyond money} \cite{bellotti2015motivation, kuhn2017micro}.

{Despite the importance of such values in the maker entrepreneurs' businesses, our interviews suggest that they are often hard to track.} 
The consequence of this is that traditional small business tools focused on tracking assets, liabilities, profit, and loss fail to help maker entrepreneurs track and reason about what they value. 
From this perspective tools, training, and mentorship for maker entrepreneurs need to take seriously the "intangibles" that traditional sales and accounting platforms and business courses overlook. 
For example, {online marketplaces for maker entrepreneurs could emphasize years of experience and time put into making a good rather than money spent on the materials when helping someone consider pricing. }
Accounting systems could be developed that help maker entrepreneurs track their enjoyment during creative exploration or the feelings they have when developing products alongside tracking costs and revenues. {More specifically, an app that helps maker entrepreneurs prioritize time and money could have a feature that allows users to custom name a unit of currency important to them, like creative freedom or enjoyment.}
{Furthermore, eCommerce platforms and social media tools used for marketing and sales could help makers track the value in their relationships through tracking positive community engagement or their feelings about their communication with fans alongside their sales data and view counts.}

\subsection{Limitations \& Future Work}

Because of the snowball sampling method from the researchers' personal connections, the level of education of our interviewees tended to be high, and the interviewees were only in the U.S. Because of this relatively small sample, we would need to see if the findings generalize to the larger maker entrepreneur population. 

The COVID-19 pandemic, as well as distance, limited the in-person interview opportunities, like the ability for a researcher to see and ask about a maker entrepreneur's studio processes. On the video call, some participants pointed to the materials, tools, and products in their homes, which helped us understand their environment and the materials and tools they had at their disposal. 

Most maker entrepreneurs did not use online marketplaces, like Etsy or Amazon Handmade, even though these are highly popular. Thus, our results may be biased by people who prefer to run everything on their own, but they also shed light on a subset of the maker entrepreneur community that online marketplaces may not serve well at this time. More research is required to find out what proportion of maker entrepreneurs use online marketplaces versus a personal assemblage of platforms. 

Because of our literature analysis on making internationally, we were aware of cultural differences that could have come up within the U.S.. For example, we kept an eye out for differences between entrepreneurial training in different genders, as described by \citet{afroze2014women}, but did not find a big divide. Future work could include a larger sample size to see if there are significant differences in this respect. 

%% file: 05_conclusion.tex
\section{Conclusion}
Our interview study with 20 maker entrepreneurs, six creative entrepreneurs, and seven people who support creative entrepreneurs 

highlights the tension these makers feel between their creative passions and the practical challenges of entrepreneurship. 
Their needs go beyond that of other similarly-sized small businesses, in large part because of the way that the creative aspects of their work affect their identity and values.
These differences have important ramifications for technologies and training to support this group.
Our research highlights a need for more tailored solutions in the context of maker entrepreneurs' small-scale, physical creative goods-oriented businesses. 
Based on our findings, we provide recommendations for designers to consider in their development of tools to provide business support for maker entrepreneurs, focusing on supporting the unique identity, values, and training needs of maker entrepreneurs.
Overall, we intend for this work to contribute to the intersection of HCI and entrepreneurship and to the development of new digital tools to help maker entrepreneurs operate successful businesses.

%% file: Appendix.tex
\newpage
\section*{Appendix A: Semi-structured Interview Questions}
\label{appendixA}
\begin{table}[h]
\caption{General interview guide for conducting semi-structured interviews.}
\label{tab:interview}
\begin{tabular}{|l|l|}
\hline
\textbf{Category}                                             & \textbf{Question}                                                                                  \\ \hline
\multirow{3}{*}{\textbf{Introduction and Background}}         & Can you introduce yourself?                                                                        \\ \cline{2-2} 
                                                              & Can you describe what motivated you?                                                               \\ \cline{2-2} 
 &
  \begin{tabular}[c]{@{}l@{}}Can you describe how they got started, \\ practically and financially?\end{tabular} \\ \hline
\multirow{5}{*}{\textbf{The business promotion questions}}    & How do you promote your business?                                                                  \\ \cline{2-2} 
                                                              & Who are your target customers?                                                                     \\ \cline{2-2} 
                                                              & How did you reach your first customers?                                                            \\ \cline{2-2} 
                                                              & \begin{tabular}[c]{@{}l@{}}What’s common about your customer’s \\ shopping behaviors?\end{tabular} \\ \cline{2-2} 
                                                              & What do you think your selling points are?                                                         \\ \hline
\multirow{4}{*}{\textbf{The factory \& production questions}} & How do you produce your products?                                                                  \\ \cline{2-2} 
 &
  \begin{tabular}[c]{@{}l@{}}What challenges have you encountered \\ in manufacturing? (If working with a factory)\end{tabular} \\ \cline{2-2} 
                                                              & How did you find the factory for production?                                                       \\ \cline{2-2} 
                                                              & Are you satisfied with the production timeline?                                                    \\ \hline
\multirow{7}{*}{\textbf{The artistic story questions}} &
  \begin{tabular}[c]{@{}l@{}}Do you convey your personal stories and \\ creative intentions to customers?\end{tabular} \\ \cline{2-2} 
                                                              & Do you think it’s helpful for your business?                                                       \\ \cline{2-2} 
                                                              & What form/platform?                                                                                \\ \cline{2-2} 
                                                              & Do you wish for a better tool for telling stories?                                                 \\ \cline{2-2} 
                                                              & Why are customers interested in the stories?                                                       \\ \cline{2-2} 
                                                              & Considering the time commitment, is it worth it?                                                   \\ \cline{2-2} 
                                                              & Do you enjoy conveying personal stories?                                                           \\ \hline
\multirow{6}{*}{\textbf{Reflection questions}}                & What other challenges did you face?                                                                \\ \cline{2-2} 
                                                              & How did you overcome the challenges?                                                               \\ \cline{2-2} 
                                                              & What’s the cause of the challenges?                                                                \\ \cline{2-2} 
 &
  \begin{tabular}[c]{@{}l@{}}What do you enjoy the most about \\ being a creative entrepreneur?\end{tabular} \\ \cline{2-2} 
 &
  \begin{tabular}[c]{@{}l@{}}What is something you wish you \\ had known earlier in starting your business?\end{tabular} \\ \cline{2-2} 
                                                              & What does success look like to you?                                                                \\ \hline
\end{tabular}

\end{table}

%% file: 00_main.bbl

 \newcommand{\noop}[1]{}
\begin{thebibliography}{102}


\ifx \showCODEN    \undefined \def \showCODEN     #1{\unskip}     \fi
\ifx \showISBNx    \undefined \def \showISBNx     #1{\unskip}     \fi
\ifx \showISBNxiii \undefined \def \showISBNxiii  #1{\unskip}     \fi
\ifx \showISSN     \undefined \def \showISSN      #1{\unskip}     \fi
\ifx \showLCCN     \undefined \def \showLCCN      #1{\unskip}     \fi
\ifx \shownote     \undefined \def \shownote      #1{#1}          \fi
\ifx \showarticletitle \undefined \def \showarticletitle #1{#1}   \fi
\ifx \showURL      \undefined \def \showURL       {\relax}        \fi
\providecommand\bibfield[2]{#2}
\providecommand\bibinfo[2]{#2}
\providecommand\natexlab[1]{#1}
\providecommand\showeprint[2][]{arXiv:#2}

\bibitem[Acker and Murthy(2018)]%
        {acker_venmo_2018}
\bibfield{author}{\bibinfo{person}{Amelia Acker} {and} \bibinfo{person}{Dhiraj Murthy}.} \bibinfo{year}{2018}\natexlab{}.
\newblock \showarticletitle{Venmo: {Understanding} {Mobile} {Payments} as {Social} {Media}}. In \bibinfo{booktitle}{\emph{Proceedings of the 9th {International} {Conference} on {Social} {Media} and {Society}}} \emph{(\bibinfo{series}{{SMSociety} '18})}. \bibinfo{publisher}{Association for Computing Machinery}, \bibinfo{address}{Copenhagen, Denmark}, \bibinfo{pages}{5--12}.
\newblock
\showISBNx{978-1-4503-6334-1}
\href{https://doi.org/10.1145/3217804.3217892}{doi:\nolinkurl{10.1145/3217804.3217892}}


\bibitem[Afroze et~al\mbox{.}(2014)]%
        {afroze2014women}
\bibfield{author}{\bibinfo{person}{Tania Afroze}, \bibinfo{person}{Md~Kashrul Alam}, \bibinfo{person}{Eliza Akther}, {and} \bibinfo{person}{Nahid~Sultana Jui}.} \bibinfo{year}{2014}\natexlab{}.
\newblock \showarticletitle{Women entrepreneurs in Bangladesh-Challenges and determining factors}.
\newblock \bibinfo{journal}{\emph{Journal of Business and Technology (Dhaka)}} \bibinfo{volume}{9}, \bibinfo{number}{2} (\bibinfo{year}{2014}), \bibinfo{pages}{27--41}.
\newblock


\bibitem[Aldrich(2014)]%
        {aldrich_democratization_2014}
\bibfield{author}{\bibinfo{person}{Howard~E. Aldrich}.} \bibinfo{year}{2014}\natexlab{}.
\newblock \showarticletitle{The democratization of entrepreneurship? {Hackers}, makerspaces, and crowdfunding}.
\newblock \bibinfo{journal}{\emph{Academy of Management Proceedings}} \bibinfo{volume}{2014}, \bibinfo{number}{1} (\bibinfo{year}{2014}), \bibinfo{pages}{10622}.
\newblock


\bibitem[Aldrich and Kenworthy(1999)]%
        {aldrich1999accidental}
\bibfield{author}{\bibinfo{person}{Howard~E Aldrich} {and} \bibinfo{person}{Amy Kenworthy}.} \bibinfo{year}{1999}\natexlab{}.
\newblock \showarticletitle{The accidental entrepreneur: Campbellian antinomies and organizational foundings}.
\newblock \bibinfo{journal}{\emph{Variations in organization science: In honor of Donald T. Campbell}} (\bibinfo{year}{1999}), \bibinfo{pages}{19--33}.
\newblock


\bibitem[Anderson(2012)]%
        {anderson_makers_2012}
\bibfield{author}{\bibinfo{person}{Chris Anderson}.} \bibinfo{year}{2012}\natexlab{}.
\newblock \bibinfo{booktitle}{\emph{Makers: {The} {New} {Industrial} {Revolution}}}.
\newblock \bibinfo{publisher}{Crown}.
\newblock
\showISBNx{978-0-307-72097-9}


\bibitem[Annett(2020)]%
        {homepreneurship}
\bibfield{author}{\bibinfo{person}{Michelle Annett}.} \bibinfo{year}{2020}\natexlab{}.
\newblock \showarticletitle{Understanding the Homepreneurship Opportunities Afforded by Social Networking and Personal Fabrication Technologies}.
\newblock \bibinfo{journal}{\emph{Proc. ACM Hum.-Comput. Interact.}} \bibinfo{volume}{4}, \bibinfo{number}{CSCW2}, Article \bibinfo{articleno}{99} (\bibinfo{date}{oct} \bibinfo{year}{2020}), \bibinfo{numpages}{48}~pages.
\newblock
\href{https://doi.org/10.1145/3415170}{doi:\nolinkurl{10.1145/3415170}}


\bibitem[Avle et~al\mbox{.}(2019)]%
        {avle2019additional}
\bibfield{author}{\bibinfo{person}{Seyram Avle}, \bibinfo{person}{Julie Hui}, \bibinfo{person}{Silvia Lindtner}, {and} \bibinfo{person}{Tawanna Dillahunt}.} \bibinfo{year}{2019}\natexlab{}.
\newblock \showarticletitle{Additional labors of the entrepreneurial self}.
\newblock \bibinfo{journal}{\emph{Proceedings of the ACM on Human-Computer Interaction}} \bibinfo{volume}{3}, \bibinfo{number}{CSCW} (\bibinfo{year}{2019}), \bibinfo{pages}{1--24}.
\newblock


\bibitem[Avle et~al\mbox{.}(2017)]%
        {avle_how_2017}
\bibfield{author}{\bibinfo{person}{Seyram Avle}, \bibinfo{person}{Silvia Lindtner}, {and} \bibinfo{person}{Kaiton Williams}.} \bibinfo{year}{2017}\natexlab{}.
\newblock \showarticletitle{How {Methods} {Make} {Designers}}. In \bibinfo{booktitle}{\emph{Proceedings of the 2017 {CHI} {Conference} on {Human} {Factors} in {Computing} {Systems}}} \emph{(\bibinfo{series}{{CHI} '17})}. \bibinfo{publisher}{ACM}, \bibinfo{address}{New York, NY, USA}, \bibinfo{pages}{472--483}.
\newblock
\showISBNx{978-1-4503-4655-9}
\href{https://doi.org/10.1145/3025453.3025864}{doi:\nolinkurl{10.1145/3025453.3025864}}


\bibitem[Banks(2010)]%
        {banks_autonomy_2010}
\bibfield{author}{\bibinfo{person}{Mark Banks}.} \bibinfo{year}{2010}\natexlab{}.
\newblock \showarticletitle{Autonomy {Guaranteed}? {Cultural} {Work} and the “{Art}–{Commerce} {Relation}”}.
\newblock \bibinfo{journal}{\emph{Journal for Cultural Research}} \bibinfo{volume}{14}, \bibinfo{number}{3} (\bibinfo{date}{July} \bibinfo{year}{2010}), \bibinfo{pages}{251--269}.
\newblock
\showISSN{1479-7585, 1740-1666}
\href{https://doi.org/10.1080/14797581003791487}{doi:\nolinkurl{10.1080/14797581003791487}}


\bibitem[Bellotti et~al\mbox{.}(2015)]%
        {bellotti2015motivation}
\bibfield{author}{\bibinfo{person}{Victoria Bellotti}, \bibinfo{person}{Alexander Ambard}, \bibinfo{person}{Daniel Turner}, \bibinfo{person}{Christina Gossmann}, \bibinfo{person}{Kamila Demkova}, {and} \bibinfo{person}{John~M. Carroll}.} \bibinfo{year}{2015}\natexlab{}.
\newblock \showarticletitle{A Muddle of Models of Motivation for Using Peer-to-Peer Economy Systems}. In \bibinfo{booktitle}{\emph{Proceedings of the 33rd Annual ACM Conference on Human Factors in Computing Systems}} (Seoul, Republic of Korea) \emph{(\bibinfo{series}{CHI '15})}. \bibinfo{publisher}{Association for Computing Machinery}, \bibinfo{address}{New York, NY, USA}, \bibinfo{pages}{1085–1094}.
\newblock
\showISBNx{9781450331456}
\href{https://doi.org/10.1145/2702123.2702272}{doi:\nolinkurl{10.1145/2702123.2702272}}


\bibitem[Berglund et~al\mbox{.}(2020)]%
        {berglund_opportunities_2020}
\bibfield{author}{\bibinfo{person}{Henrik Berglund}, \bibinfo{person}{Marouane Bousfiha}, {and} \bibinfo{person}{Yashar Mansoori}.} \bibinfo{year}{2020}\natexlab{}.
\newblock \showarticletitle{Opportunities as {Artifacts} and {Entrepreneurship} as {Design}.}
\newblock \bibinfo{journal}{\emph{Academy of Management Review}} (\bibinfo{date}{Feb.} \bibinfo{year}{2020}).
\newblock
\showISSN{0363-7425}
\href{https://doi.org/10.5465/amr.2018.0285}{doi:\nolinkurl{10.5465/amr.2018.0285}}
\newblock
\shownote{Publisher: Academy of Management}.


\bibitem[Bhansing et~al\mbox{.}(2018)]%
        {bhansing2018passion}
\bibfield{author}{\bibinfo{person}{Pawan~V Bhansing}, \bibinfo{person}{Erik Hitters}, {and} \bibinfo{person}{Yosha Wijngaarden}.} \bibinfo{year}{2018}\natexlab{}.
\newblock \showarticletitle{Passion inspires: Motivations of creative entrepreneurs in creative business centres in the Netherlands}.
\newblock \bibinfo{journal}{\emph{The Journal of Entrepreneurship}} \bibinfo{volume}{27}, \bibinfo{number}{1} (\bibinfo{year}{2018}), \bibinfo{pages}{1--24}.
\newblock


\bibitem[Browder et~al\mbox{.}(2019)]%
        {browder_emergence_2019}
\bibfield{author}{\bibinfo{person}{Russell~E. Browder}, \bibinfo{person}{Howard~E. Aldrich}, {and} \bibinfo{person}{Steven~W. Bradley}.} \bibinfo{year}{2019}\natexlab{}.
\newblock \showarticletitle{The emergence of the maker movement: {Implications} for entrepreneurship research}.
\newblock \bibinfo{journal}{\emph{Journal of Business Venturing}} \bibinfo{volume}{34}, \bibinfo{number}{3} (\bibinfo{date}{May} \bibinfo{year}{2019}), \bibinfo{pages}{459--476}.
\newblock
\showISSN{0883-9026}
\href{https://doi.org/10.1016/j.jbusvent.2019.01.005}{doi:\nolinkurl{10.1016/j.jbusvent.2019.01.005}}


\bibitem[Casey(2014)]%
        {casey2014critical}
\bibfield{author}{\bibinfo{person}{Colleen Casey}.} \bibinfo{year}{2014}\natexlab{}.
\newblock \showarticletitle{Critical connections: The importance of community-based organizations and social capital to credit access for low-wealth entrepreneurs}.
\newblock \bibinfo{journal}{\emph{Urban Affairs Review}} \bibinfo{volume}{50}, \bibinfo{number}{3} (\bibinfo{year}{2014}), \bibinfo{pages}{366--390}.
\newblock


\bibitem[Crain and Bailey(2017)]%
        {crain2017share}
\bibfield{author}{\bibinfo{person}{Patrick~A Crain} {and} \bibinfo{person}{Brian~P Bailey}.} \bibinfo{year}{2017}\natexlab{}.
\newblock \showarticletitle{Share Once or Share Often: Exploring How Designers Approach Iteration in a Large Online Community}. In \bibinfo{booktitle}{\emph{Proceedings of the 2017 ACM SIGCHI Conference on Creativity and Cognition}}. ACM, \bibinfo{publisher}{Sheridan Communications}, \bibinfo{address}{Singapore}, \bibinfo{pages}{80--92}.
\newblock


\bibitem[Dabbish et~al\mbox{.}(2012)]%
        {dabbish2012social}
\bibfield{author}{\bibinfo{person}{Laura Dabbish}, \bibinfo{person}{Colleen Stuart}, \bibinfo{person}{Jason Tsay}, {and} \bibinfo{person}{Jim Herbsleb}.} \bibinfo{year}{2012}\natexlab{}.
\newblock \showarticletitle{Social coding in GitHub: transparency and collaboration in an open software repository}. In \bibinfo{booktitle}{\emph{Proceedings of the ACM 2012 conference on computer supported cooperative work}}. \bibinfo{pages}{1277--1286}.
\newblock


\bibitem[DeFillippi et~al\mbox{.}(2007)]%
        {defillippi_introduction_2007}
\bibfield{author}{\bibinfo{person}{Robert DeFillippi}, \bibinfo{person}{Gernot Grabher}, {and} \bibinfo{person}{Candace Jones}.} \bibinfo{year}{2007}\natexlab{}.
\newblock \showarticletitle{Introduction to paradoxes of creativity: managerial and organizational challenges in the cultural economy}.
\newblock \bibinfo{journal}{\emph{Journal of Organizational Behavior}} \bibinfo{volume}{28}, \bibinfo{number}{5} (\bibinfo{date}{July} \bibinfo{year}{2007}), \bibinfo{pages}{511--521}.
\newblock
\showISSN{08943796, 10991379}
\href{https://doi.org/10.1002/job.466}{doi:\nolinkurl{10.1002/job.466}}


\bibitem[Deresiewicz(2015)]%
        {deresiewicz2015death}
\bibfield{author}{\bibinfo{person}{William Deresiewicz}.} \bibinfo{year}{2015}\natexlab{}.
\newblock \showarticletitle{The Death of the Artist—and the Birth of the Creative Entrepreneur}.
\newblock \bibinfo{journal}{\emph{The Atlantic}} \bibinfo{volume}{315}, \bibinfo{number}{1} (\bibinfo{year}{2015}), \bibinfo{pages}{92}.
\newblock


\bibitem[Dillahunt et~al\mbox{.}(2018)]%
        {dillahunt2018entrepreneurship}
\bibfield{author}{\bibinfo{person}{Tawanna~R Dillahunt}, \bibinfo{person}{Vaishnav Kameswaran}, \bibinfo{person}{Desiree McLain}, \bibinfo{person}{Minnie Lester}, \bibinfo{person}{Delores Orr}, {and} \bibinfo{person}{Kentaro Toyama}.} \bibinfo{year}{2018}\natexlab{}.
\newblock \showarticletitle{Entrepreneurship and the socio-technical chasm in a lean economy}. In \bibinfo{booktitle}{\emph{Proceedings of the 2018 CHI Conference on Human Factors in Computing Systems}}. \bibinfo{pages}{1--14}.
\newblock


\bibitem[Dillahunt et~al\mbox{.}(2022)]%
        {dillahunt2022village}
\bibfield{author}{\bibinfo{person}{Tawanna~R Dillahunt}, \bibinfo{person}{Alex~Jiahong Lu}, \bibinfo{person}{Aarti Israni}, \bibinfo{person}{Ruchita Lodha}, \bibinfo{person}{Savana Brewer}, \bibinfo{person}{Tiera~S Robinson}, \bibinfo{person}{Angela~Brown Wilson}, {and} \bibinfo{person}{Earnest Wheeler}.} \bibinfo{year}{2022}\natexlab{}.
\newblock \showarticletitle{The Village: Infrastructuring Community-based Mentoring to Support Adults Experiencing Poverty}. In \bibinfo{booktitle}{\emph{Proceedings of the 2022 CHI Conference on Human Factors in Computing Systems}}. \bibinfo{pages}{1--17}.
\newblock


\bibitem[Dougherty(2012)]%
        {dougherty_maker_2012}
\bibfield{author}{\bibinfo{person}{Dale Dougherty}.} \bibinfo{year}{2012}\natexlab{}.
\newblock \showarticletitle{The {Maker} {Movement}}.
\newblock \bibinfo{journal}{\emph{Innovations: Technology, Governance, Globalization}} \bibinfo{volume}{7}, \bibinfo{number}{3} (\bibinfo{date}{July} \bibinfo{year}{2012}), \bibinfo{pages}{11--14}.
\newblock
\showISSN{1558-2477, 1558-2485}
\href{https://doi.org/10.1162/INOV_a_00135}{doi:\nolinkurl{10.1162/INOV_a_00135}}


\bibitem[Doussard et~al\mbox{.}(2018)]%
        {doussard_manufacturing_2018}
\bibfield{author}{\bibinfo{person}{Marc Doussard}, \bibinfo{person}{Greg Schrock}, \bibinfo{person}{Laura Wolf-Powers}, \bibinfo{person}{Max Eisenburger}, {and} \bibinfo{person}{Stephen Marotta}.} \bibinfo{year}{2018}\natexlab{}.
\newblock \showarticletitle{Manufacturing without the firm: {Challenges} for the maker movement in three {U}.{S}. cities}.
\newblock \bibinfo{journal}{\emph{Environment and Planning A: Economy and Space}} \bibinfo{volume}{50}, \bibinfo{number}{3} (\bibinfo{date}{May} \bibinfo{year}{2018}), \bibinfo{pages}{651--670}.
\newblock
\showISSN{0308-518X}
\href{https://doi.org/10.1177/0308518X17749709}{doi:\nolinkurl{10.1177/0308518X17749709}}


\bibitem[Duffy(2017)]%
        {duffy2017not}
\bibfield{author}{\bibinfo{person}{Brooke~Erin Duffy}.} \bibinfo{year}{2017}\natexlab{}.
\newblock \bibinfo{booktitle}{\emph{(Not) getting paid to do what you love: Gender, social media, and aspirational work}}.
\newblock \bibinfo{publisher}{Yale University Press}.
\newblock


\bibitem[Duffy(2018)]%
        {duffy2018}
\bibfield{author}{\bibinfo{person}{Brooke~Erin Duffy}.} \bibinfo{year}{2018}\natexlab{}.
\newblock \showarticletitle{Entrepreneurial Wishes and Career Dreams}.
\newblock In \bibinfo{booktitle}{\emph{(Not) Getting Paid to Do What You Love}}. \bibinfo{publisher}{Yale University Press}, \bibinfo{pages}{1--11}.
\newblock


\bibitem[Duffy and Pruchniewska(2017)]%
        {duffy2017gender}
\bibfield{author}{\bibinfo{person}{Brooke~Erin Duffy} {and} \bibinfo{person}{Urszula Pruchniewska}.} \bibinfo{year}{2017}\natexlab{}.
\newblock \showarticletitle{Gender and self-enterprise in the social media age: A digital double bind}.
\newblock \bibinfo{journal}{\emph{Information, communication \& society}} \bibinfo{volume}{20}, \bibinfo{number}{6} (\bibinfo{year}{2017}), \bibinfo{pages}{843--859}.
\newblock


\bibitem[Eikhof and Haunschild(2007)]%
        {eikhof_for_2007}
\bibfield{author}{\bibinfo{person}{Doris~Ruth Eikhof} {and} \bibinfo{person}{Axel Haunschild}.} \bibinfo{year}{2007}\natexlab{}.
\newblock \showarticletitle{For art's sake! {Artistic} and economic logics in creative production}.
\newblock \bibinfo{journal}{\emph{Journal of Organizational Behavior}} \bibinfo{volume}{28}, \bibinfo{number}{5} (\bibinfo{date}{July} \bibinfo{year}{2007}), \bibinfo{pages}{523--538}.
\newblock
\showISSN{08943796, 10991379}
\href{https://doi.org/10.1002/job.462}{doi:\nolinkurl{10.1002/job.462}}


\bibitem[Essig(2017)]%
        {essig2017same}
\bibfield{author}{\bibinfo{person}{Linda Essig}.} \bibinfo{year}{2017}\natexlab{}.
\newblock \showarticletitle{Same or different? The “cultural entrepreneurship” and “arts entrepreneurship” constructs in European and US higher education}.
\newblock \bibinfo{journal}{\emph{Cultural Trends}} \bibinfo{volume}{26}, \bibinfo{number}{2} (\bibinfo{year}{2017}), \bibinfo{pages}{125--137}.
\newblock


\bibitem[Etsy(2020)]%
        {etsy_sell_nodate}
\bibfield{author}{\bibinfo{person}{Etsy}.} \bibinfo{year}{2020}\natexlab{}.
\newblock \bibinfo{title}{The {Sell} on {Etsy} {App} for {iPhone} and {iPad}}.
\newblock
\urldef\tempurl%
\url{http://help.etsy.com/hc/en-us/articles/115015711428-The-Sell-on-Etsy-App-for-iPhone-and-iPad}
\showURL{%
\tempurl}


\bibitem[{Etsy Inc.}(2019)]%
        {etsytransparency2019}
\bibfield{author}{\bibinfo{person}{{Etsy Inc.}}} \bibinfo{year}{2019}\natexlab{}.
\newblock \bibinfo{title}{Etys 2019 Transparency Report}.
\newblock
\urldef\tempurl%
\url{https://extfiles.etsy.com/advocacy/Etsy\%5F2019\%5FTransparency\%5FReport.pdf}
\showURL{%
\tempurl}


\bibitem[Etsy.com(2017)]%
        {etsywomen}
\bibfield{author}{\bibinfo{person}{Etsy.com}.} \bibinfo{year}{2017}\natexlab{}.
\newblock \bibinfo{title}{Etys 2017 Impact Report}.
\newblock \bibinfo{howpublished}{\url{https://extfiles.etsy.com/Impact/2017EtsyImpactUpdate.pdf}}.
\newblock
\newblock
\shownote{Accessed: 2020-01-23}.


\bibitem[Etsy.com(2021)]%
        {etsy2021}
\bibfield{author}{\bibinfo{person}{Etsy.com}.} \bibinfo{year}{2021}\natexlab{}.
\newblock \bibinfo{title}{Etys 2021 Impact Report}.
\newblock \bibinfo{howpublished}{\url{https://storage.googleapis.com/etsy-extfiles-prod/Press/reports/2021_Etsy_Transparency_Report.pdf?ref=news}}.
\newblock
\newblock
\shownote{Accessed: 2023-01-09}.


\bibitem[Foong et~al\mbox{.}(2017)]%
        {foong2017online}
\bibfield{author}{\bibinfo{person}{Eureka Foong}, \bibinfo{person}{Steven~P Dow}, \bibinfo{person}{Brian~P Bailey}, {and} \bibinfo{person}{Elizabeth~M Gerber}.} \bibinfo{year}{2017}\natexlab{}.
\newblock \showarticletitle{Online feedback exchange: A framework for understanding the socio-psychological factors}. In \bibinfo{booktitle}{\emph{Proceedings of the 2017 CHI Conference on Human Factors in Computing Systems}}. ACM, \bibinfo{publisher}{Sheridan Communications}, \bibinfo{address}{Denver, Colorado}, \bibinfo{pages}{4454--4467}.
\newblock


\bibitem[Garvin et~al\mbox{.}(2023)]%
        {garvin2023counter}
\bibfield{author}{\bibinfo{person}{Matthew Garvin}, \bibinfo{person}{Ron Eglash}, \bibinfo{person}{Kwame~Porter Robinson}, \bibinfo{person}{Lionel Robert}, \bibinfo{person}{Mark Guzdial}, {and} \bibinfo{person}{Audrey Bennett}.} \bibinfo{year}{2023}\natexlab{}.
\newblock \showarticletitle{Counter-hegemonic AI: The Role of Artisanal Identity in the Design of Automation for a Liberated Economy}.
\newblock In \bibinfo{booktitle}{\emph{AI and the Future of Creative Work}}. \bibinfo{publisher}{Routledge}, \bibinfo{pages}{71--88}.
\newblock


\bibitem[Gerber and Hui(2013)]%
        {gerber2013crowdfunding}
\bibfield{author}{\bibinfo{person}{Elizabeth~M Gerber} {and} \bibinfo{person}{Julie Hui}.} \bibinfo{year}{2013}\natexlab{}.
\newblock \showarticletitle{Crowdfunding: Motivations and deterrents for participation}.
\newblock \bibinfo{journal}{\emph{ACM Transactions on Computer-Human Interaction (TOCHI)}} \bibinfo{volume}{20}, \bibinfo{number}{6} (\bibinfo{year}{2013}), \bibinfo{pages}{1--32}.
\newblock


\bibitem[Glăveanu and Tanggaard(2014)]%
        {glaveanu_creativity_2014}
\bibfield{author}{\bibinfo{person}{Vlad~Petre Glăveanu} {and} \bibinfo{person}{Lene Tanggaard}.} \bibinfo{year}{2014}\natexlab{}.
\newblock \showarticletitle{Creativity, identity, and representation: {Towards} a socio-cultural theory of creative identity}.
\newblock \bibinfo{journal}{\emph{New Ideas in Psychology}} \bibinfo{volume}{34}, \bibinfo{number}{1} (\bibinfo{year}{2014}), \bibinfo{pages}{12--21}.
\newblock
\showISSN{0732-118X}
\href{https://doi.org/10.1016/j.newideapsych.2014.02.002}{doi:\nolinkurl{10.1016/j.newideapsych.2014.02.002}}
\newblock
\shownote{Publisher: Elsevier Ltd}.


\bibitem[Gotsi et~al\mbox{.}(2010)]%
        {gotsi_managing_2010}
\bibfield{author}{\bibinfo{person}{Manto Gotsi}, \bibinfo{person}{Constantine Andriopoulos}, \bibinfo{person}{Marianne~W Lewis}, {and} \bibinfo{person}{Amy~E Ingram}.} \bibinfo{year}{2010}\natexlab{}.
\newblock \showarticletitle{Managing creatives: {Paradoxical} approaches to identity regulation}.
\newblock \bibinfo{journal}{\emph{Human Relations}} \bibinfo{volume}{63}, \bibinfo{number}{6} (\bibinfo{date}{June} \bibinfo{year}{2010}), \bibinfo{pages}{781--805}.
\newblock
\showISSN{0018-7267, 1741-282X}
\href{https://doi.org/10.1177/0018726709342929}{doi:\nolinkurl{10.1177/0018726709342929}}


\bibitem[Harboe and Huang(2015)]%
        {harboe2015real}
\bibfield{author}{\bibinfo{person}{Gunnar Harboe} {and} \bibinfo{person}{Elaine~M Huang}.} \bibinfo{year}{2015}\natexlab{}.
\newblock \showarticletitle{Real-world affinity diagramming practices: Bridging the paper-digital gap}. In \bibinfo{booktitle}{\emph{Proceedings of the 33rd annual ACM conference on human factors in computing systems}}. \bibinfo{pages}{95--104}.
\newblock


\bibitem[Hedditch and Vyas(2023)]%
        {womenIntersectional}
\bibfield{author}{\bibinfo{person}{Sonali Hedditch} {and} \bibinfo{person}{Dhaval Vyas}.} \bibinfo{year}{2023}\natexlab{}.
\newblock \showarticletitle{Crossing the Threshold: Pathways into Makerspaces for Women at the Intersectional Margins}.
\newblock \bibinfo{journal}{\emph{Proc. ACM Hum.-Comput. Interact.}} \bibinfo{volume}{7}, \bibinfo{number}{CSCW1}, Article \bibinfo{articleno}{123} (\bibinfo{date}{apr} \bibinfo{year}{2023}), \bibinfo{numpages}{40}~pages.
\newblock
\href{https://doi.org/10.1145/3579599}{doi:\nolinkurl{10.1145/3579599}}


\bibitem[Hennekam and Bennett(2016)]%
        {hennekam_involuntary_2016}
\bibfield{author}{\bibinfo{person}{Sophie Hennekam} {and} \bibinfo{person}{Dawn Bennett}.} \bibinfo{year}{2016}\natexlab{}.
\newblock \showarticletitle{Involuntary career transition and identity within the artist population}.
\newblock \bibinfo{journal}{\emph{Personnel Review}} \bibinfo{volume}{45}, \bibinfo{number}{6} (\bibinfo{date}{Sept.} \bibinfo{year}{2016}), \bibinfo{pages}{1114--1131}.
\newblock
\showISSN{0048-3486}
\href{https://doi.org/10.1108/PR-01-2015-0020}{doi:\nolinkurl{10.1108/PR-01-2015-0020}}


\bibitem[Hisrich et~al\mbox{.}(2017)]%
        {hisrich2017entrepreneurship}
\bibfield{author}{\bibinfo{person}{Robert~D Hisrich}, \bibinfo{person}{Michael~P Peters}, {and} \bibinfo{person}{Dean~A Shepherd}.} \bibinfo{year}{2017}\natexlab{}.
\newblock \bibinfo{booktitle}{\emph{Entrepreneurship}}.
\newblock \bibinfo{publisher}{McGraw-Hill Education}.
\newblock


\bibitem[Howard et~al\mbox{.}(2014)]%
        {howard2014maker}
\bibfield{author}{\bibinfo{person}{Charles Howard}, \bibinfo{person}{Andrea Gerosa}, \bibinfo{person}{Maria~Carrasco Mejuto}, {and} \bibinfo{person}{Gregor Giannella}.} \bibinfo{year}{2014}\natexlab{}.
\newblock \showarticletitle{The Maker Movement: a new avenue for competition in the EU}.
\newblock \bibinfo{journal}{\emph{European View}} \bibinfo{volume}{13}, \bibinfo{number}{2} (\bibinfo{year}{2014}), \bibinfo{pages}{333--340}.
\newblock


\bibitem[Hui et~al\mbox{.}(2020)]%
        {hui2020community}
\bibfield{author}{\bibinfo{person}{Julie Hui}, \bibinfo{person}{Nefer~Ra Barber}, \bibinfo{person}{Wendy Casey}, \bibinfo{person}{Suzanne Cleage}, \bibinfo{person}{Danny~C Dolley}, \bibinfo{person}{Frances Worthy}, \bibinfo{person}{Kentaro Toyama}, {and} \bibinfo{person}{Tawanna~R Dillahunt}.} \bibinfo{year}{2020}\natexlab{}.
\newblock \showarticletitle{Community collectives: Low-tech social support for digitally-engaged entrepreneurship}. In \bibinfo{booktitle}{\emph{Proceedings of the 2020 CHI conference on human factors in computing systems}}. \bibinfo{pages}{1--15}.
\newblock


\bibitem[Hui et~al\mbox{.}(2018)]%
        {hui2018making}
\bibfield{author}{\bibinfo{person}{Julie Hui}, \bibinfo{person}{Kentaro Toyama}, \bibinfo{person}{Joyojeet Pal}, {and} \bibinfo{person}{Tawanna Dillahunt}.} \bibinfo{year}{2018}\natexlab{}.
\newblock \showarticletitle{Making a living my way: Necessity-driven entrepreneurship in resource-constrained communities}.
\newblock \bibinfo{journal}{\emph{Proceedings of the ACM on Human-Computer Interaction}} \bibinfo{volume}{2}, \bibinfo{number}{CSCW} (\bibinfo{year}{2018}), \bibinfo{pages}{1--24}.
\newblock


\bibitem[Hui et~al\mbox{.}(2019)]%
        {hui2019distributed}
\bibfield{author}{\bibinfo{person}{Julie~S Hui}, \bibinfo{person}{Matthew~W Easterday}, {and} \bibinfo{person}{Elizabeth~M Gerber}.} \bibinfo{year}{2019}\natexlab{}.
\newblock \showarticletitle{Distributed apprenticeship in online communities}.
\newblock \bibinfo{journal}{\emph{Human--Computer Interaction}} \bibinfo{volume}{34}, \bibinfo{number}{4} (\bibinfo{year}{2019}), \bibinfo{pages}{328--378}.
\newblock


\bibitem[Hui and Gerber(2017)]%
        {hui2017makerspaces-entrepreneurship}
\bibfield{author}{\bibinfo{person}{Julie~S. Hui} {and} \bibinfo{person}{Elizabeth~M. Gerber}.} \bibinfo{year}{2017}\natexlab{}.
\newblock \showarticletitle{Developing Makerspaces as Sites of Entrepreneurship}. In \bibinfo{booktitle}{\emph{Proceedings of the 2017 ACM Conference on Computer Supported Cooperative Work and Social Computing}} (Portland, Oregon, USA) \emph{(\bibinfo{series}{CSCW '17})}. \bibinfo{publisher}{Association for Computing Machinery}, \bibinfo{address}{New York, NY, USA}, \bibinfo{pages}{2023–2038}.
\newblock
\showISBNx{9781450343350}
\href{https://doi.org/10.1145/2998181.2998264}{doi:\nolinkurl{10.1145/2998181.2998264}}


\bibitem[Israni et~al\mbox{.}(2023)]%
        {israni2023opportunities}
\bibfield{author}{\bibinfo{person}{Aarti Israni}, \bibinfo{person}{Julie Hui}, {and} \bibinfo{person}{Tawanna~R Dillahunt}.} \bibinfo{year}{2023}\natexlab{}.
\newblock \showarticletitle{Opportunities for Social Media to Support Aspiring Entrepreneurs with Financial Constraints}.
\newblock \bibinfo{journal}{\emph{Proceedings of the ACM on Human-Computer Interaction}} \bibinfo{volume}{7}, \bibinfo{number}{CSCW1} (\bibinfo{year}{2023}), \bibinfo{pages}{1--27}.
\newblock


\bibitem[Khaire(2017)]%
        {khaire_culture_2017}
\bibfield{author}{\bibinfo{person}{Mukti Khaire}.} \bibinfo{year}{2017}\natexlab{}.
\newblock \bibinfo{booktitle}{\emph{Culture and {Commerce}: {The} {Value} of {Entrepreneurship} in {Creative} {Industries}}}.
\newblock \bibinfo{publisher}{Stanford University Press}.
\newblock
\showISBNx{978-1-5036-0308-0}


\bibitem[Khaire(2021)]%
        {khaire2021entrepreneurship}
\bibfield{author}{\bibinfo{person}{Mukti Khaire}.} \bibinfo{year}{2021}\natexlab{}.
\newblock \showarticletitle{Entrepreneurship by design: The construction of meanings and markets for cultural craft goods}.
\newblock In \bibinfo{booktitle}{\emph{Culture, Innovation and Entrepreneurship}}. \bibinfo{publisher}{Routledge}, \bibinfo{pages}{13--32}.
\newblock


\bibitem[Kim et~al\mbox{.}(2017)]%
        {kim2017mosaic}
\bibfield{author}{\bibinfo{person}{Joy Kim}, \bibinfo{person}{Maneesh Agrawala}, {and} \bibinfo{person}{Michael~S Bernstein}.} \bibinfo{year}{2017}\natexlab{}.
\newblock \showarticletitle{Mosaic: designing online creative communities for sharing works-in-progress}. In \bibinfo{booktitle}{\emph{Proceedings of the 2017 ACM Conference on Computer Supported Cooperative Work and Social Computing}}. ACM, \bibinfo{publisher}{Sheridan Communications}, \bibinfo{address}{Portland, Oregon}, \bibinfo{pages}{246--258}.
\newblock


\bibitem[Klawitter(2017)]%
        {klawitter2017independent}
\bibfield{author}{\bibinfo{person}{Erin~Flynn Klawitter}.} \bibinfo{year}{2017}\natexlab{}.
\newblock \emph{\bibinfo{title}{How Independent Artists Participate in the Peer Economy for Handmade Goods}}.
\newblock \bibinfo{thesistype}{Ph.\,D. Dissertation}. \bibinfo{school}{Northwestern University}.
\newblock


\bibitem[Kokkalis et~al\mbox{.}(2017)]%
        {socialcapital}
\bibfield{author}{\bibinfo{person}{Nicolas Kokkalis}, \bibinfo{person}{Chengdiao Fan}, \bibinfo{person}{Thomas Breier}, {and} \bibinfo{person}{Michael~S. Bernstein}.} \bibinfo{year}{2017}\natexlab{}.
\newblock \showarticletitle{Founder Center: Enabling Access to Collective Social Capital}. In \bibinfo{booktitle}{\emph{Proceedings of the 2017 ACM Conference on Computer Supported Cooperative Work and Social Computing}} (Portland, Oregon, USA) \emph{(\bibinfo{series}{CSCW '17})}. \bibinfo{publisher}{Association for Computing Machinery}, \bibinfo{address}{New York, NY, USA}, \bibinfo{pages}{2010–2022}.
\newblock
\showISBNx{9781450343350}
\href{https://doi.org/10.1145/2998181.2998244}{doi:\nolinkurl{10.1145/2998181.2998244}}


\bibitem[Kotturi et~al\mbox{.}(2021)]%
        {kotturi2021unique}
\bibfield{author}{\bibinfo{person}{Yasmine Kotturi}, \bibinfo{person}{Allie Blaising}, \bibinfo{person}{Sarah~E Fox}, {and} \bibinfo{person}{Chinmay Kulkarni}.} \bibinfo{year}{2021}\natexlab{}.
\newblock \showarticletitle{The Unique Challenges for Creative Small Businesses Seeking Feedback on Social Media}.
\newblock \bibinfo{journal}{\emph{Proceedings of the ACM on Human-Computer Interaction}} \bibinfo{volume}{5}, \bibinfo{number}{CSCW1} (\bibinfo{year}{2021}), \bibinfo{pages}{1--27}.
\newblock


\bibitem[Kotturi et~al\mbox{.}(2022)]%
        {kotturi2022tech}
\bibfield{author}{\bibinfo{person}{Yasmine Kotturi}, \bibinfo{person}{Herman~T Johnson}, \bibinfo{person}{Michael Skirpan}, \bibinfo{person}{Sarah~E Fox}, \bibinfo{person}{Jeffrey~P Bigham}, {and} \bibinfo{person}{Amy Pavel}.} \bibinfo{year}{2022}\natexlab{}.
\newblock \showarticletitle{Tech Help Desk: Support for Local Entrepreneurs Addressing the Long Tail of Computing Challenges}. In \bibinfo{booktitle}{\emph{CHI Conference on Human Factors in Computing Systems}}. \bibinfo{pages}{1--15}.
\newblock


\bibitem[Kotturi et~al\mbox{.}(2024)]%
        {kotturi2024peerdea}
\bibfield{author}{\bibinfo{person}{Yasmine Kotturi}, \bibinfo{person}{Jenny Yu}, \bibinfo{person}{Pranav Khadpe}, \bibinfo{person}{Erin Gatz}, \bibinfo{person}{Harvey Zheng}, \bibinfo{person}{Sarah~E Fox}, {and} \bibinfo{person}{Chinmay Kulkarni}.} \bibinfo{year}{2024}\natexlab{}.
\newblock \showarticletitle{Peerdea: Co-Designing a Peer Support Platform with Creative Entrepreneurs}.
\newblock \bibinfo{journal}{\emph{Proceedings of the ACM on Human-Computer Interaction}} \bibinfo{number}{CSCW1} (\bibinfo{year}{2024}), \bibinfo{pages}{1--27}.
\newblock


\bibitem[Kreiner et~al\mbox{.}(2006)]%
        {kreiner_where_2006}
\bibfield{author}{\bibinfo{person}{Glen Kreiner}, \bibinfo{person}{Elaine Hollensbe}, {and} \bibinfo{person}{Mathew Sheep}.} \bibinfo{year}{2006}\natexlab{}.
\newblock \showarticletitle{WHERE IS THE ``ME'' AMONG THE ``WE''? IDENTITY WORK AND THE SEARCH FOR OPTIMAL BALANCE}.
\newblock \bibinfo{journal}{\emph{Academy of Management Journal}} \bibinfo{volume}{49}, \bibinfo{number}{5} (\bibinfo{year}{2006}), \bibinfo{pages}{1031--1057}.
\newblock
\showISSN{0001-4273}
\href{https://doi.org/10.5465/AMJ.2006.22798186}{doi:\nolinkurl{10.5465/AMJ.2006.22798186}}


\bibitem[Krishna~Kumaran et~al\mbox{.}(2021)]%
        {krishna2021plan}
\bibfield{author}{\bibinfo{person}{Sneha~R Krishna~Kumaran}, \bibinfo{person}{Yue Yin}, {and} \bibinfo{person}{Brian~P Bailey}.} \bibinfo{year}{2021}\natexlab{}.
\newblock \showarticletitle{Plan early, revise more: effects of goal setting and perceived role of the feedback provider on feedback seeking behavior}.
\newblock \bibinfo{journal}{\emph{Proceedings of the ACM on Human-Computer Interaction}} \bibinfo{volume}{5}, \bibinfo{number}{CSCW1} (\bibinfo{year}{2021}), \bibinfo{pages}{1--22}.
\newblock


\bibitem[Kuhn et~al\mbox{.}(2016)]%
        {kuhn2016near}
\bibfield{author}{\bibinfo{person}{Kristine Kuhn}, \bibinfo{person}{Tera Galloway}, {and} \bibinfo{person}{Maureen Collins-Williams}.} \bibinfo{year}{2016}\natexlab{}.
\newblock \showarticletitle{Near, far, and online: Small business owners’ advice-seeking from peers}.
\newblock \bibinfo{journal}{\emph{Journal of Small Business and Enterprise Development}} \bibinfo{volume}{23}, \bibinfo{number}{1} (\bibinfo{year}{2016}), \bibinfo{pages}{189--206}.
\newblock


\bibitem[Kuhn and Maleki(2017)]%
        {kuhn2017micro}
\bibfield{author}{\bibinfo{person}{Kristine~M Kuhn} {and} \bibinfo{person}{Amir Maleki}.} \bibinfo{year}{2017}\natexlab{}.
\newblock \showarticletitle{Micro-entrepreneurs, dependent contractors, and instaserfs: Understanding online labor platform workforces}.
\newblock \bibinfo{journal}{\emph{Academy of Management Perspectives}} \bibinfo{volume}{31}, \bibinfo{number}{3} (\bibinfo{year}{2017}), \bibinfo{pages}{183--200}.
\newblock


\bibitem[Lang(2013)]%
        {lang_zero_2013}
\bibfield{author}{\bibinfo{person}{David Lang}.} \bibinfo{year}{2013}\natexlab{}.
\newblock \bibinfo{booktitle}{\emph{Zero to {M}aker: {L}earn (just enough) to make (just about) anything} (\bibinfo{edition}{first edition.} ed.)}.
\newblock \bibinfo{publisher}{Maker Media}, \bibinfo{address}{Sebastopol, California}.
\newblock
\showISBNx{978-1-4493-5641-5}


\bibitem[Lazar et~al\mbox{.}(2021)]%
        {10.1145/3411764.3445146}
\bibfield{author}{\bibinfo{person}{Amanda Lazar}, \bibinfo{person}{Alisha Pradhan}, \bibinfo{person}{Ben Jelen}, \bibinfo{person}{Katie A.~Siek}, {and} \bibinfo{person}{Alex Leitch}.} \bibinfo{year}{2021}\natexlab{}.
\newblock \showarticletitle{Studying the Formation of an Older Adult-Led Makerspace}. In \bibinfo{booktitle}{\emph{Proceedings of the 2021 CHI Conference on Human Factors in Computing Systems}} (Yokohama, Japan) \emph{(\bibinfo{series}{CHI '21})}. \bibinfo{publisher}{Association for Computing Machinery}, \bibinfo{address}{New York, NY, USA}, Article \bibinfo{articleno}{593}, \bibinfo{numpages}{11}~pages.
\newblock
\showISBNx{9781450380966}
\href{https://doi.org/10.1145/3411764.3445146}{doi:\nolinkurl{10.1145/3411764.3445146}}


\bibitem[Lee et~al\mbox{.}(2023)]%
        {lee2023refugee}
\bibfield{author}{\bibinfo{person}{Chuike Lee}, \bibinfo{person}{Stephen Viller}, {and} \bibinfo{person}{Dhaval Vyas}.} \bibinfo{year}{2023}\natexlab{}.
\newblock \showarticletitle{Refugee Entrepreneurial Trajectories}.
\newblock \bibinfo{journal}{\emph{Proceedings of the ACM on Human-Computer Interaction}} \bibinfo{volume}{7}, \bibinfo{number}{CSCW2} (\bibinfo{year}{2023}), \bibinfo{pages}{1--26}.
\newblock


\bibitem[Lindtner(2015)]%
        {lindtner2015hacking}
\bibfield{author}{\bibinfo{person}{Silvia Lindtner}.} \bibinfo{year}{2015}\natexlab{}.
\newblock \showarticletitle{Hacking with Chinese characteristics: The promises of the maker movement against China’s manufacturing culture}.
\newblock \bibinfo{journal}{\emph{Science, Technology, \& Human Values}} \bibinfo{volume}{40}, \bibinfo{number}{5} (\bibinfo{year}{2015}), \bibinfo{pages}{854--879}.
\newblock


\bibitem[Lindtner et~al\mbox{.}(2015)]%
        {lindtner2015designed}
\bibfield{author}{\bibinfo{person}{Silvia Lindtner}, \bibinfo{person}{Anna Greenspan}, {and} \bibinfo{person}{David Li}.} \bibinfo{year}{2015}\natexlab{}.
\newblock \showarticletitle{Designed in Shenzhen: Shanzhai manufacturing and maker entrepreneurs}. In \bibinfo{booktitle}{\emph{Proceedings of the fifth decennial Aarhus conference on critical alternatives}}. \bibinfo{pages}{85--96}.
\newblock


\bibitem[Lindtner et~al\mbox{.}(2014)]%
        {lindtner2014emerging}
\bibfield{author}{\bibinfo{person}{Silvia Lindtner}, \bibinfo{person}{Garnet~D Hertz}, {and} \bibinfo{person}{Paul Dourish}.} \bibinfo{year}{2014}\natexlab{}.
\newblock \showarticletitle{Emerging sites of HCI innovation: hackerspaces, hardware startups \& incubators}. In \bibinfo{booktitle}{\emph{Proceedings of the SIGCHI conference on human factors in computing systems}}. \bibinfo{pages}{439--448}.
\newblock


\bibitem[Lipson and Kurman(2010)]%
        {lipson2010factory}
\bibfield{author}{\bibinfo{person}{Hod Lipson} {and} \bibinfo{person}{Melba Kurman}.} \bibinfo{year}{2010}\natexlab{}.
\newblock \showarticletitle{Factory@ home: The emerging economy of personal fabrication}.
\newblock \bibinfo{journal}{\emph{A report commissioned by the US Office of Science and Technology Policy}} (\bibinfo{year}{2010}).
\newblock


\bibitem[Ljungblad(2023)]%
        {ljungblad2023applying}
\bibfield{author}{\bibinfo{person}{Sara Ljungblad}.} \bibinfo{year}{2023}\natexlab{}.
\newblock \showarticletitle{Applying “Designerly Framing” to Understand Assisted Feeding as Social Aesthetic Bodily Experiences}.
\newblock \bibinfo{journal}{\emph{ACM Transactions on Human-Robot Interaction}} \bibinfo{volume}{12}, \bibinfo{number}{2} (\bibinfo{year}{2023}), \bibinfo{pages}{1--23}.
\newblock


\bibitem[Luckman(2013)]%
        {luckman_aura_2013}
\bibfield{author}{\bibinfo{person}{Susan Luckman}.} \bibinfo{year}{2013}\natexlab{}.
\newblock \showarticletitle{The {Aura} of the {Analogue} in a {Digital} {Age}: {Women}’s {Crafts}, {Creative} {Markets} and {Home}-{Based} {Labour} {After} {Etsy}}.
\newblock \bibinfo{journal}{\emph{Cultural Studies Review}} \bibinfo{volume}{19}, \bibinfo{number}{1} (\bibinfo{date}{Feb.} \bibinfo{year}{2013}), \bibinfo{pages}{249--70--249--70}.
\newblock
\showISSN{1837-8692}
\href{https://doi.org/10.5130/csr.v19i1.2585}{doi:\nolinkurl{10.5130/csr.v19i1.2585}}
\newblock
\shownote{Number: 1}.


\bibitem[Makridis and Kuuskoski(2023)]%
        {makridis2023narrowing}
\bibfield{author}{\bibinfo{person}{Christos Makridis} {and} \bibinfo{person}{Jonathan Kuuskoski}.} \bibinfo{year}{2023}\natexlab{}.
\newblock \showarticletitle{Narrowing the Gap: Implications of Arts Business Training on Artist Labor Market Outcomes}.
\newblock \bibinfo{journal}{\emph{Artivate: A Journal of Entrepreneurship in the Arts}}  \bibinfo{volume}{12} (\bibinfo{year}{2023}).
\newblock


\bibitem[Markman and Baron(2003)]%
        {markman2003person}
\bibfield{author}{\bibinfo{person}{Gideon~D Markman} {and} \bibinfo{person}{Robert~A Baron}.} \bibinfo{year}{2003}\natexlab{}.
\newblock \showarticletitle{Person--entrepreneurship fit: why some people are more successful as entrepreneurs than others}.
\newblock \bibinfo{journal}{\emph{Human resource management review}} \bibinfo{volume}{13}, \bibinfo{number}{2} (\bibinfo{year}{2003}), \bibinfo{pages}{281--301}.
\newblock


\bibitem[McDonald et~al\mbox{.}(2019)]%
        {mcdonald2019reliability}
\bibfield{author}{\bibinfo{person}{Nora McDonald}, \bibinfo{person}{Sarita Schoenebeck}, {and} \bibinfo{person}{Andrea Forte}.} \bibinfo{year}{2019}\natexlab{}.
\newblock \showarticletitle{Reliability and inter-rater reliability in qualitative research: Norms and guidelines for CSCW and HCI practice}.
\newblock \bibinfo{journal}{\emph{Proceedings of the ACM on human-computer interaction}} \bibinfo{volume}{3}, \bibinfo{number}{CSCW} (\bibinfo{year}{2019}), \bibinfo{pages}{1--23}.
\newblock


\bibitem[McRobbie(2002)]%
        {mcrobbie_clubs_2002}
\bibfield{author}{\bibinfo{person}{Angela McRobbie}.} \bibinfo{year}{2002}\natexlab{}.
\newblock \showarticletitle{CLUBS TO COMPANIES: NOTES ON THE DECLINE OF POLITICAL CULTURE IN SPEEDED UP CREATIVE WORLDS}.
\newblock \bibinfo{journal}{\emph{Cultural Studies}} \bibinfo{volume}{16}, \bibinfo{number}{4} (\bibinfo{date}{July} \bibinfo{year}{2002}), \bibinfo{pages}{516--531}.
\newblock
\showISSN{0950-2386, 1466-4348}
\href{https://doi.org/10.1080/09502380210139098}{doi:\nolinkurl{10.1080/09502380210139098}}


\bibitem[Meera and Vinodan(2022)]%
        {meera2022innovative}
\bibfield{author}{\bibinfo{person}{S Meera} {and} \bibinfo{person}{A Vinodan}.} \bibinfo{year}{2022}\natexlab{}.
\newblock \showarticletitle{Innovative approach and marketing skill: a case study of artisan entrepreneurs of India}.
\newblock \bibinfo{journal}{\emph{Journal of Entrepreneurship in Emerging Economies}} (\bibinfo{year}{2022}).
\newblock


\bibitem[Michlewski(2008)]%
        {michlewski_uncovering_2008}
\bibfield{author}{\bibinfo{person}{Kamil Michlewski}.} \bibinfo{year}{2008}\natexlab{}.
\newblock \showarticletitle{Uncovering {Design} {Attitude}: {Inside} the {Culture} of {Designers}}.
\newblock \bibinfo{journal}{\emph{Organization Studies}} \bibinfo{volume}{29}, \bibinfo{number}{3} (\bibinfo{date}{March} \bibinfo{year}{2008}), \bibinfo{pages}{373--392}.
\newblock
\showISSN{0170-8406, 1741-3044}
\href{https://doi.org/10.1177/0170840607088019}{doi:\nolinkurl{10.1177/0170840607088019}}


\bibitem[Miller(2007)]%
        {miller_etsy_2007}
\bibfield{author}{\bibinfo{person}{Kerry Miller}.} \bibinfo{year}{2007}\natexlab{}.
\newblock \showarticletitle{Etsy: {A} site for artisans takes off}.
\newblock \bibinfo{journal}{\emph{Business Week June}}  \bibinfo{volume}{12} (\bibinfo{year}{2007}).
\newblock


\bibitem[Naderifar et~al\mbox{.}(2017)]%
        {naderifar2017snowball}
\bibfield{author}{\bibinfo{person}{Mahin Naderifar}, \bibinfo{person}{Hamideh Goli}, {and} \bibinfo{person}{Fereshteh Ghaljaie}.} \bibinfo{year}{2017}\natexlab{}.
\newblock \showarticletitle{Snowball sampling: A purposeful method of sampling in qualitative research}.
\newblock \bibinfo{journal}{\emph{Strides in development of medical education}} \bibinfo{volume}{14}, \bibinfo{number}{3} (\bibinfo{year}{2017}).
\newblock


\bibitem[Okerlund et~al\mbox{.}(2021)]%
        {10.1145/3411764.3445126}
\bibfield{author}{\bibinfo{person}{Johanna Okerlund}, \bibinfo{person}{David Wilson}, {and} \bibinfo{person}{Celine Latulipe}.} \bibinfo{year}{2021}\natexlab{}.
\newblock \showarticletitle{A Feminist Utopian Perspective on the Practice and Promise of Making}. In \bibinfo{booktitle}{\emph{Proceedings of the 2021 CHI Conference on Human Factors in Computing Systems}} (Yokohama, Japan) \emph{(\bibinfo{series}{CHI '21})}. \bibinfo{publisher}{Association for Computing Machinery}, \bibinfo{address}{New York, NY, USA}, Article \bibinfo{articleno}{402}, \bibinfo{numpages}{16}~pages.
\newblock
\showISBNx{9781450380966}
\href{https://doi.org/10.1145/3411764.3445126}{doi:\nolinkurl{10.1145/3411764.3445126}}


\bibitem[Olshan(2017)]%
        {olshan_after_2017}
\bibfield{author}{\bibinfo{person}{Kelly Olshan}.} \bibinfo{year}{2017}\natexlab{}.
\newblock \showarticletitle{After {Art} {School}: {Professional} {Development} {Training} in {Nonprofit} {Organizations}}.
\newblock \bibinfo{journal}{\emph{The Journal of Arts Management, Law, and Society}} \bibinfo{volume}{47}, \bibinfo{number}{4} (\bibinfo{date}{Aug.} \bibinfo{year}{2017}), \bibinfo{pages}{230--244}.
\newblock
\showISSN{1063-2921, 1930-7799}
\href{https://doi.org/10.1080/10632921.2017.1340210}{doi:\nolinkurl{10.1080/10632921.2017.1340210}}


\bibitem[OpenStax(2020)]%
        {openstax_entrepreneurship_2020}
\bibfield{author}{\bibinfo{person}{OpenStax}.} \bibinfo{year}{2020}\natexlab{}.
\newblock \bibinfo{booktitle}{\emph{Entrepreneurship by {OpenStax}} (\bibinfo{edition}{first edition} ed.)}.
\newblock \bibinfo{publisher}{XanEdu Publishing Inc}, \bibinfo{address}{Houston, Texas}.
\newblock
\showISBNx{978-1-975076-34-4}


\bibitem[Poorsoltan(2012)]%
        {poorsoltan_artists_2012}
\bibfield{author}{\bibinfo{person}{Keramat Poorsoltan}.} \bibinfo{year}{2012}\natexlab{}.
\newblock \showarticletitle{Artists as entrepreneurs}.
\newblock \bibinfo{journal}{\emph{International Journal of Entrepreneurship}}  \bibinfo{volume}{16} (\bibinfo{year}{2012}), \bibinfo{pages}{83}.
\newblock
\showISSN{1099-9264}
\newblock
\shownote{Publisher: The DreamCatchers Group, LLC}.


\bibitem[Pop-Cohu{\c{t}}(2019)]%
        {pop2019online}
\bibfield{author}{\bibinfo{person}{Ioana~Crina Pop-Cohu{\c{t}}}.} \bibinfo{year}{2019}\natexlab{}.
\newblock \showarticletitle{Online Platforms and the Simulation of Creative Entrepreneurship—The Role of Creativity and the Perception about Work and Self-Employment}.
\newblock \bibinfo{journal}{\emph{LIMEN 2019}} (\bibinfo{year}{2019}), \bibinfo{pages}{113}.
\newblock


\bibitem[Pret and Cogan(2019)]%
        {pret2019artisan}
\bibfield{author}{\bibinfo{person}{Tobias Pret} {and} \bibinfo{person}{Aviel Cogan}.} \bibinfo{year}{2019}\natexlab{}.
\newblock \showarticletitle{Artisan entrepreneurship: a systematic literature review and research agenda}.
\newblock \bibinfo{journal}{\emph{International Journal of Entrepreneurial Behavior \& Research}} \bibinfo{volume}{25}, \bibinfo{number}{4} (\bibinfo{year}{2019}), \bibinfo{pages}{592--614}.
\newblock


\bibitem[Rao(2023)]%
        {rao2023transcribing}
\bibfield{author}{\bibinfo{person}{Ashwin Rao}.} \bibinfo{year}{2023}\natexlab{}.
\newblock \showarticletitle{Transcribing Educational Videos Using Whisper: A preliminary study on using AI for transcribing educational videos}.
\newblock \bibinfo{journal}{\emph{arXiv preprint arXiv:2307.03200}} (\bibinfo{year}{2023}).
\newblock


\bibitem[Rosner et~al\mbox{.}(2014)]%
        {rosner2014making}
\bibfield{author}{\bibinfo{person}{Daniela~K Rosner}, \bibinfo{person}{Silvia Lindtner}, \bibinfo{person}{Ingrid Erickson}, \bibinfo{person}{Laura Forlano}, \bibinfo{person}{Steven~J Jackson}, {and} \bibinfo{person}{Beth Kolko}.} \bibinfo{year}{2014}\natexlab{}.
\newblock \showarticletitle{Making cultures: building things \& building communities}. In \bibinfo{booktitle}{\emph{Proceedings of the companion publication of the 17th ACM conference on Computer supported cooperative work \& social computing}}. \bibinfo{pages}{113--116}.
\newblock


\bibitem[Sahut et~al\mbox{.}(2021)]%
        {sahut2021age}
\bibfield{author}{\bibinfo{person}{Jean-Michel Sahut}, \bibinfo{person}{Luca Iandoli}, {and} \bibinfo{person}{Fr{\'e}d{\'e}ric Teulon}.} \bibinfo{year}{2021}\natexlab{}.
\newblock \showarticletitle{The age of digital entrepreneurship}.
\newblock \bibinfo{journal}{\emph{Small Business Economics}}  \bibinfo{volume}{56} (\bibinfo{year}{2021}), \bibinfo{pages}{1159--1169}.
\newblock


\bibitem[Shultz(2015)]%
        {shultz_work_2015}
\bibfield{author}{\bibinfo{person}{Benjamin Shultz}.} \bibinfo{year}{2015}\natexlab{}.
\newblock \showarticletitle{The {Work} {Behind} the {Scenes}: {The} {New} {Intermediaries} of the {Indie} {Crafts} {Business}}.
\newblock \bibinfo{journal}{\emph{Regional Studies}} \bibinfo{volume}{49}, \bibinfo{number}{3} (\bibinfo{date}{March} \bibinfo{year}{2015}).
\newblock
\showISSN{0034-3404}
\href{https://doi.org/10.1080/00343404.2013.770597}{doi:\nolinkurl{10.1080/00343404.2013.770597}}


\bibitem[Skaggs et~al\mbox{.}(2017)]%
        {skaggs2017strategic}
\bibfield{author}{\bibinfo{person}{R Skaggs}, \bibinfo{person}{A Frenette}, \bibinfo{person}{S Gaskill}, {and} \bibinfo{person}{AL Miller}.} \bibinfo{year}{2017}\natexlab{}.
\newblock \showarticletitle{{Strategic National Arts Alumni Project-2017 Annual Report: Arts Alumni in their Communities}}.
\newblock \bibinfo{journal}{\emph{Annual Report}} (\bibinfo{year}{2017}).
\newblock
\newblock
\shownote{http://snaap.indiana.edu/pdf/2017/SNAAP\_Annual\_Report\_2017.pdf}.


\bibitem[Solomon and Mathias(2020)]%
        {solomon2020artisans}
\bibfield{author}{\bibinfo{person}{Shelby~J Solomon} {and} \bibinfo{person}{Blake~D Mathias}.} \bibinfo{year}{2020}\natexlab{}.
\newblock \showarticletitle{The artisans' dilemma: Artisan entrepreneurship and the challenge of firm growth}.
\newblock \bibinfo{journal}{\emph{Journal of Business Venturing}} \bibinfo{volume}{35}, \bibinfo{number}{5} (\bibinfo{year}{2020}), \bibinfo{pages}{106044}.
\newblock


\bibitem[Sullivan(2000)]%
        {sullivan2000entrepreneurial}
\bibfield{author}{\bibinfo{person}{Robert Sullivan}.} \bibinfo{year}{2000}\natexlab{}.
\newblock \showarticletitle{Entrepreneurial learning and mentoring}.
\newblock \bibinfo{journal}{\emph{International Journal of Entrepreneurial Behavior \& Research}} \bibinfo{volume}{6}, \bibinfo{number}{3} (\bibinfo{year}{2000}), \bibinfo{pages}{160--175}.
\newblock


\bibitem[Sultan and Sharmin(2020)]%
        {sultan2020exploratory}
\bibfield{author}{\bibinfo{person}{Mohammad~Tipu Sultan} {and} \bibinfo{person}{Farzana Sharmin}.} \bibinfo{year}{2020}\natexlab{}.
\newblock \showarticletitle{An exploratory investigation of facebook live marketing by women entrepreneurs in bangladesh}. In \bibinfo{booktitle}{\emph{Social Computing and Social Media. Participation, User Experience, Consumer Experience, and Applications of Social Computing: 12th International Conference, SCSM 2020, Held as Part of the 22nd HCI International Conference, HCII 2020, Copenhagen, Denmark, July 19--24, 2020, Proceedings, Part II 22}}. Springer, \bibinfo{pages}{415--430}.
\newblock


\bibitem[Tajfel and Turner(1979)]%
        {tajfel_integrative_1979}
\bibfield{author}{\bibinfo{person}{Henri Tajfel} {and} \bibinfo{person}{John~C. Turner}.} \bibinfo{year}{1979}\natexlab{}.
\newblock \showarticletitle{An integrative theory of intergroup conflict}.
\newblock \bibinfo{journal}{\emph{Organizational identity: A reader}}  \bibinfo{volume}{56} (\bibinfo{year}{1979}), \bibinfo{pages}{65}.
\newblock
\newblock
\shownote{Publisher: Oxford University Press Oxford, UK}.


\bibitem[Tan and Ming(2003)]%
        {tan2003leveraging}
\bibfield{author}{\bibinfo{person}{T Tan} {and} \bibinfo{person}{Matthew Ming}.} \bibinfo{year}{2003}\natexlab{}.
\newblock \showarticletitle{Leveraging on symbolic values and meanings in branding}.
\newblock \bibinfo{journal}{\emph{Journal of brand Management}}  \bibinfo{volume}{10} (\bibinfo{year}{2003}), \bibinfo{pages}{208--218}.
\newblock


\bibitem[Tracy and Trethewey(2006)]%
        {tracy_fracturing_2006}
\bibfield{author}{\bibinfo{person}{Sarah~J. Tracy} {and} \bibinfo{person}{Angela Trethewey}.} \bibinfo{year}{2006}\natexlab{}.
\newblock \showarticletitle{Fracturing the {Real}-{Self}$\leftrightarrow${Fake}-{Self} {Dichotomy}: {Moving} {Toward} “{Crystallized}” {Organizational} {Discourses} and {Identities}}.
\newblock \bibinfo{journal}{\emph{Communication Theory}} \bibinfo{volume}{15}, \bibinfo{number}{2} (\bibinfo{date}{Jan.} \bibinfo{year}{2006}), \bibinfo{pages}{168--195}.
\newblock
\showISSN{1050-3293}
\href{https://doi.org/10.1111/j.1468-2885.2005.tb00331.x}{doi:\nolinkurl{10.1111/j.1468-2885.2005.tb00331.x}}


\bibitem[Troxler and Wolf(2017)]%
        {troxler2017digital}
\bibfield{author}{\bibinfo{person}{Peter Troxler} {and} \bibinfo{person}{Patricia Wolf}.} \bibinfo{year}{2017}\natexlab{}.
\newblock \showarticletitle{Digital maker-entrepreneurs in open design: What activities make up their business model?}
\newblock \bibinfo{journal}{\emph{Business horizons}} \bibinfo{volume}{60}, \bibinfo{number}{6} (\bibinfo{year}{2017}), \bibinfo{pages}{807--817}.
\newblock


\bibitem[Von~Hippel(2016)]%
        {von2016free}
\bibfield{author}{\bibinfo{person}{Eric Von~Hippel}.} \bibinfo{year}{2016}\natexlab{}.
\newblock \bibinfo{booktitle}{\emph{Free innovation}}.
\newblock \bibinfo{publisher}{MIT press}.
\newblock


\bibitem[Vyas(2020)]%
        {DIY}
\bibfield{author}{\bibinfo{person}{Dhaval Vyas}.} \bibinfo{year}{2020}\natexlab{}.
\newblock \showarticletitle{Life Improvements: DIY in Low Socio-economic Status Communities}. In \bibinfo{booktitle}{\emph{Companion Publication of the 2020 Conference on Computer Supported Cooperative Work and Social Computing}} (Virtual Event, USA) \emph{(\bibinfo{series}{CSCW '20 Companion})}. \bibinfo{publisher}{Association for Computing Machinery}, \bibinfo{address}{New York, NY, USA}, \bibinfo{pages}{407–412}.
\newblock
\showISBNx{9781450380591}
\href{https://doi.org/10.1145/3406865.3418325}{doi:\nolinkurl{10.1145/3406865.3418325}}


\bibitem[Vyas and Vines(2019)]%
        {DIY2}
\bibfield{author}{\bibinfo{person}{Dhaval Vyas} {and} \bibinfo{person}{John Vines}.} \bibinfo{year}{2019}\natexlab{}.
\newblock \showarticletitle{Making at the Margins: Making in an Under-resourced e-Waste Recycling Center}.
\newblock \bibinfo{journal}{\emph{Proc. ACM Hum.-Comput. Interact.}} \bibinfo{volume}{3}, \bibinfo{number}{CSCW}, Article \bibinfo{articleno}{188} (\bibinfo{date}{nov} \bibinfo{year}{2019}), \bibinfo{numpages}{23}~pages.
\newblock
\href{https://doi.org/10.1145/3359290}{doi:\nolinkurl{10.1145/3359290}}


\bibitem[Wenger(1999)]%
        {wenger1999communities}
\bibfield{author}{\bibinfo{person}{Etienne Wenger}.} \bibinfo{year}{1999}\natexlab{}.
\newblock \bibinfo{booktitle}{\emph{Communities of practice: Learning, meaning, and identity}}.
\newblock \bibinfo{publisher}{Cambridge university press}.
\newblock


\bibitem[Werthes et~al\mbox{.}(2018)]%
        {werthes_cultural_2018}
\bibfield{author}{\bibinfo{person}{Daniela Werthes}, \bibinfo{person}{René Mauer}, {and} \bibinfo{person}{Malte Brettel}.} \bibinfo{year}{2018}\natexlab{}.
\newblock \showarticletitle{Cultural and creative entrepreneurs: understanding the role of entrepreneurial identity}.
\newblock \bibinfo{journal}{\emph{International Journal of Entrepreneurial Behavior \& Research}} \bibinfo{volume}{24}, \bibinfo{number}{1} (\bibinfo{date}{Jan.} \bibinfo{year}{2018}), \bibinfo{pages}{290--314}.
\newblock
\showISSN{1355-2554}
\href{https://doi.org/10.1108/IJEBR-07-2016-0215}{doi:\nolinkurl{10.1108/IJEBR-07-2016-0215}}


\bibitem[Williams(2021)]%
        {williams2021genz}
\bibfield{author}{\bibinfo{person}{Lara Williams}.} \bibinfo{year}{2021}\natexlab{}.
\newblock \bibinfo{title}{Gen Z Are Hustling For Their Post-Covid Futures}.
\newblock
\urldef\tempurl%
\url{https://www.bloomberg.com/opinion/articles/2021-02-26/side-hustles-help-gen-z-make-money-carve-out-a-post-covid-future}
\showURL{%
\tempurl}


\bibitem[Wolf-Powers et~al\mbox{.}(2017)]%
        {wolf2017maker}
\bibfield{author}{\bibinfo{person}{Laura Wolf-Powers}, \bibinfo{person}{Marc Doussard}, \bibinfo{person}{Greg Schrock}, \bibinfo{person}{Charles Heying}, \bibinfo{person}{Max Eisenburger}, {and} \bibinfo{person}{Stephen Marotta}.} \bibinfo{year}{2017}\natexlab{}.
\newblock \showarticletitle{The Maker Movement and Urban Economic Development}.
\newblock \bibinfo{journal}{\emph{Journal of the American Planning Association}} \bibinfo{volume}{83}, \bibinfo{number}{4} (\bibinfo{year}{2017}), \bibinfo{pages}{365--376}.
\newblock
\href{https://doi.org/10.1080/01944363.2017.1360787}{doi:\nolinkurl{10.1080/01944363.2017.1360787}}
\showeprint{https://doi.org/10.1080/01944363.2017.1360787}


\bibitem[Xu et~al\mbox{.}(2015)]%
        {xu2015classroom}
\bibfield{author}{\bibinfo{person}{Anbang Xu}, \bibinfo{person}{Huaming Rao}, \bibinfo{person}{Steven~P Dow}, {and} \bibinfo{person}{Brian~P Bailey}.} \bibinfo{year}{2015}\natexlab{}.
\newblock \showarticletitle{A classroom study of using crowd feedback in the iterative design process}. In \bibinfo{booktitle}{\emph{Proceedings of the 18th ACM conference on computer supported cooperative work \& social computing}}. \bibinfo{pages}{1637--1648}.
\newblock


\bibitem[Zhang et~al\mbox{.}(2017)]%
        {zhang_cold_2017}
\bibfield{author}{\bibinfo{person}{Xinyi Zhang}, \bibinfo{person}{Shiliang Tang}, \bibinfo{person}{Yun Zhao}, \bibinfo{person}{Gang Wang}, \bibinfo{person}{Haitao Zheng}, {and} \bibinfo{person}{Ben~Y. Zhao}.} \bibinfo{year}{2017}\natexlab{}.
\newblock \showarticletitle{Cold {Hard} {E}-{Cash}: {Friends} and {Vendors} in the {Venmo} {Digital} {Payments} {System}}. In \bibinfo{booktitle}{\emph{Eleventh {International} {AAAI} {Conference} on {Web} and {Social} {Media}}}.
\newblock
\urldef\tempurl%
\url{https://www.aaai.org/ocs/index.php/ICWSM/ICWSM17/paper/view/15589}
\showURL{%
\tempurl}


\end{thebibliography}
